%% file: rpvfinal_ppe.tex
\documentclass[12pt]{article}
\usepackage{a4wide}
\usepackage[dvips]{graphics}

\newcommand{\roots}  {\ensuremath{\sqrt{s}}}
\newcommand{\tanb} {\ensuremath{\tan \beta }}

\newcommand{\gevcc}{\mbox{GeV$/c^2$}}

\newcommand{\gev}{\mbox {GeV}}

\newcommand{\invpb}{\mbox{pb$^{-1}$}}

\newcommand{\q}{{\mathrm q}}

\newcommand{\e}{{\mathrm e}}

\newcommand{\slR}{\ensuremath{\mathrm{\tilde{\ell}_R}}}
\newcommand{\seR}{\ensuremath{\mathrm{\tilde{e}_R}}}
\newcommand{\smuR}{\ensuremath{\mathrm{\tilde{\mu}_R}}}
\newcommand{\smuL}{\ensuremath{\mathrm{\tilde{\mu}_L}}}
\newcommand{\stauR}{\ensuremath{\mathrm{\tilde{\tau}_R}}}

\newcommand{\sbL}{\ensuremath{\mathrm{\tilde{b}_L}}}
\newcommand{\stL}{\ensuremath{\mathrm{\tilde{t}_L}}}
\newcommand{\sbR}{\ensuremath{\mathrm{\tilde{b}_R}}}

\newcommand{\suL}{\ensuremath{\mathrm{\tilde{u}_L}}}

\newcommand{\sdL}{\ensuremath{\mathrm{\tilde{d}_L}}}
\newcommand{\stone}{\ensuremath{\mathrm{\tilde{t}_1}}}
\newcommand{\sbone}{\ensuremath{\mathrm{\tilde{b}_1}}}

\newcommand{\emiss}{{\not \!\! E}}

\newcommand{\ALEPH}{ALEPH}
\newcommand{\sq}{\ensuremath{\tilde{\mathrm q}}}

\newcommand{\snu}{\ensuremath{\tilde{\nu}}}
\newcommand{\snue}{\ensuremath{\snu_\mathrm{e}}}
\newcommand{\snumu}{\ensuremath{\snu_{\mu}}}
\newcommand{\slep}{\ensuremath{\tilde{\ell}}}
\newcommand{\se}{\ensuremath{\tilde{\mathrm{e}}}}

\newcommand{\stau}{\ensuremath{\tilde{\tau}}}

\newcommand{\qq}{\ensuremath{\mathrm{q\bar{q}}}}

\newcommand{\PW}{\ensuremath{\mathrm{W}}}
\newcommand{\PZ}{\ensuremath{\mathrm{Z}}}
\newcommand{\Pe}{\ensuremath{\mathrm{e}}}

\newcommand{\nch}{\ensuremath{N_{\mathrm{ch}}}}

\newcommand{\pmiss}{\ensuremath{p^\mathrm{miss}}}

\newcommand{\pzmiss}{\ensuremath{p^{\mathrm{miss}}_\mathrm{z}}}
\newcommand{\phimix}{\ensuremath{\phi_\mathrm{mix}}}

\newcommand{\newc}{\newcommand}

\newc{\R}{$R$}
\newc{\charginom}{M_{\tilde \chi}^{+}}
\newc{\mue}{\mu_{\tilde{e}_{iL}}}
\newc{\mud}{\mu_{\tilde{d}_{jL}}}
\newc{\beq}{\begin{equation}}
\newc{\eeq}{\end{equation}}
\newc{\barr}{\begin{eqnarray}}
\newc{\earr}{\end{eqnarray}}
\newc{\ra}{\rightarrow}
\newc{\Da}{\Downarrow}
\newc{\lam}{\lambda}
\newc{\eps}{\epsilon}
\newc{\eq}[1]{(\ref{eq:#1})}
\newc{\eqs}[2]{(\ref{eq:#1},\ref{eq:#2})}
\newc{\etal}{{\it et al.}\ }
\newc{\Hbar}{{\bar H}}
\newc{\Ubar}{{\bar U}}
\newc{\Dbar}{{\bar D}}
\newc{\Ebar}{{\bar E}}
\newc{\eg}{{e.g.}\ }
\newc{\ie}{{i.e.}\ }
\newc{\nonum}{\nonumber}
\newc{\lab}[1]{\label{eq:#1}}
\newc{\lle}[3]{L_{#1}L_{#2}\Ebar_{#3}}
\newc{\lqd}[3]{L_{#1}Q_{#2}\Dbar_{#3}}
\newc{\udd}[3]{\Ubar_{#1}\Dbar_{#2}\Dbar_{#3}}
\newc{\slle}{\ensuremath{LL\bar{E}}}
\newc{\slqd}{\ensuremath{LQ\bar{D}}}
\newc{\sudd}{\ensuremath{\bar{U}\bar{D}\bar{D}}}
\newc{\dpr}[2]{({#1}\cdot{#2})}
\newc{\rpv}{{\not R_p}}

\newc{\rpvm}{{\not \! R_p}}
\newc{\rp}{$R_p$}
\newc{\gsim}{ \,  \scriptstyle{\stackrel{>}{\sim}}\displaystyle \, }
\newc{\lsim}{ \,  \scriptstyle{\stackrel{<}{\sim}}\displaystyle \, }

\begin{document}
\begin{titlepage} 

\begin{center}
\rm EUROPEAN ORGANIZATION FOR NUCLEAR RESEARCH (CERN)
\end{center}

\begin{flushright}
CERN-EP/2002-071\\
%E.B. Draft\\
September 26, 2002\\
\end{flushright}

\vspace{2cm}

\begin{center} 
\LARGE{Search for Supersymmetric Particles \\ 
 with R-Parity Violating Decays \\ 
in ${\mathrm e}^+{\mathrm e}^-$ Collisions 
at $\roots$ up to 209~GeV}
\vspace{1cm}

{\large The \ALEPH\ Collaboration$^{*})$
}
\end{center}

\thispagestyle{empty}

%\maketitle

\vspace{0cm}
\begin{abstract}
\vspace{.5cm}
Searches for the pair production of supersymmetric particles under the
assumption that R-parity is violated via a single dominant $\slle$,
$\slqd$ or $\sudd$ coupling are performed using the data collected by
the \ALEPH\ detector at LEP at centre-of-mass energies from
189 to $209~\gev$.  The numbers of observed candidate events in the
data are in agreement with the Standard Model expectation, and limits
on the production cross sections and on the masses of charginos,
sleptons, squarks and sneutrinos are derived.
\end{abstract}

\vfill
\centerline{\large \em (Submitted to the European Physical Journal C) }

\begin{flushleft}
\rule{69.0mm}{0.2mm}\\
$^{*})$ See next pages for the list of authors
\end{flushleft}

\end{titlepage}

\newpage
\include{authb}

\section{Introduction}
\label{intro}
\setcounter{footnote}{0}
The search for supersymmetric particles addressed in this paper is performed
in the framework of the minimal supersymmetric extension of the Standard Model (MSSM)~\cite{MSSM}
with R-parity violation. 
Conservation of R-parity~\cite{fayet} is usually 
assumed in order to prevent experimentally 
forbidden low energy processes (e.g. fast proton decay).
Nevertheless this symmetry is not required theoretically, and models
with R-parity violation can be constructed,
which are compatible with experimental constraints.

A generic model can be built from 
the R-parity violating terms of the superpotential~\cite{rpsuper}
\begin{equation}
W_{\rpv} =  \lam_{ijk}\lle{i}{j}{k}+ \lam'_{ijk}\lqd{i}{j}{k}+
            \lam''_{ijk}\udd{i}{j}{k},
\label{eqrpv}
\end{equation}
where $\Dbar,\Ubar$ and $\Ebar$ are down-like quark, up-like quark
and lepton singlet superfields, $Q$ and $L$ are the quark and lepton
doublet superfields; $\lambda$, $\lambda'$ and $\lambda''$
are Yukawa couplings and $i,j,k=1,2,3$ are generation indices. The
presence of such R-parity violating terms implies that the lightest
supersymmetric particle (LSP) is unstable and that supersymmetric 
particles
can decay directly to Standard Model particles. 

\begin{figure}
\begin{center}
\resizebox{0.75\textwidth}{!}{
  \resizebox{0.25\textwidth}{!}{
        \includegraphics{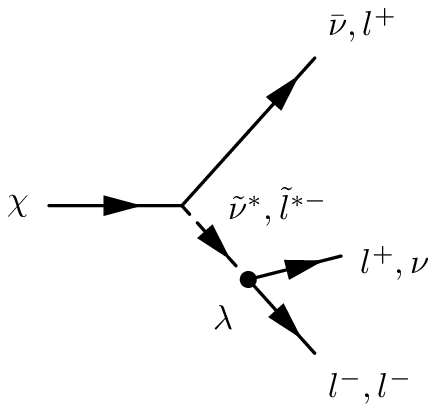}}
  \resizebox{0.25\textwidth}{!}{
        \includegraphics{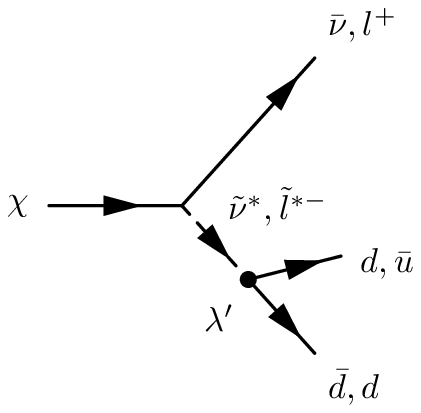}}
  \resizebox{0.25\textwidth}{!}{
        \includegraphics{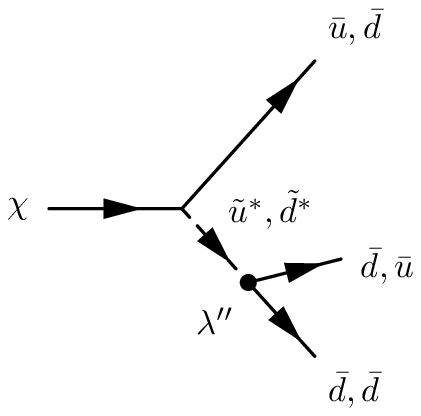}}
}
\resizebox{0.25\textwidth}{!}{
        \includegraphics{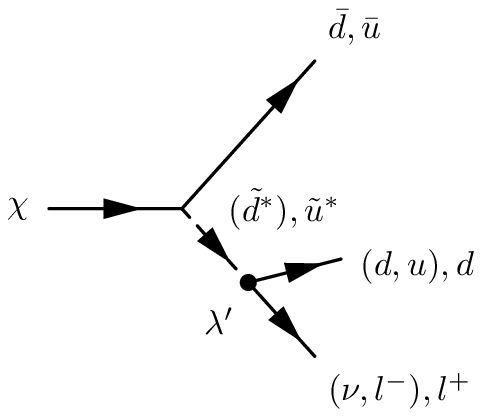}}

\resizebox{0.8\textwidth}{!}{
  \resizebox{0.25\textwidth}{!}{
        \includegraphics{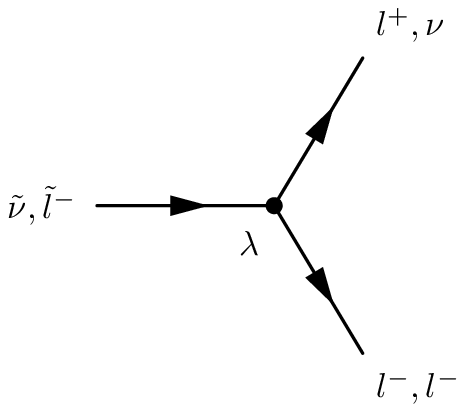}}
  \resizebox{0.25\textwidth}{!}{
        \includegraphics{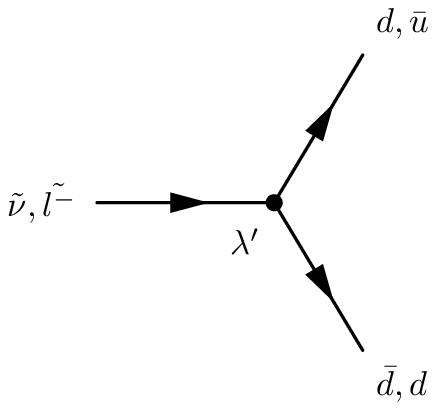}}
  \resizebox{0.25\textwidth}{!}{
        \includegraphics{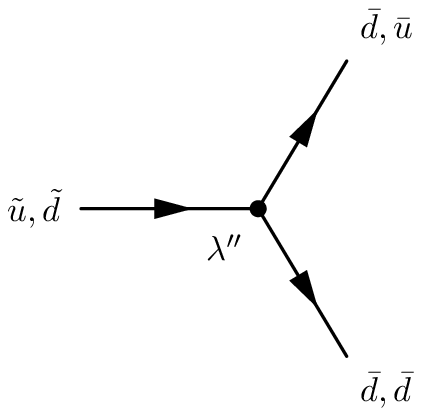}}
}
\resizebox{0.25\textwidth}{!}{
  \includegraphics{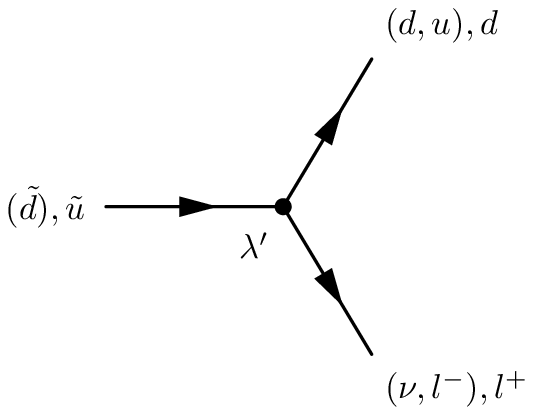}}
\end{center}
\caption[.]{ \small {\em Direct} R-parity violating decays of
supersymmetric particles via the
$\lambda$, $\lambda'$ and $\lambda''$ couplings. The points mark the
R-parity violating vertex in the decay.}
\label{dec.examples}
\end{figure}

\begin{figure}
\begin{center}
\resizebox{0.75\textwidth}{!}{
  \resizebox{0.25\textwidth}{!}{
        \includegraphics{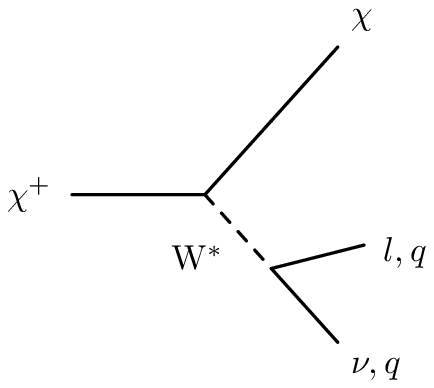}}
  \resizebox{0.25\textwidth}{!}{
        \includegraphics{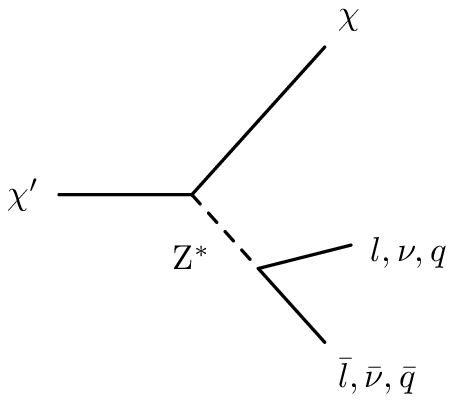}}
  \resizebox{0.25\textwidth}{!}{
        \includegraphics{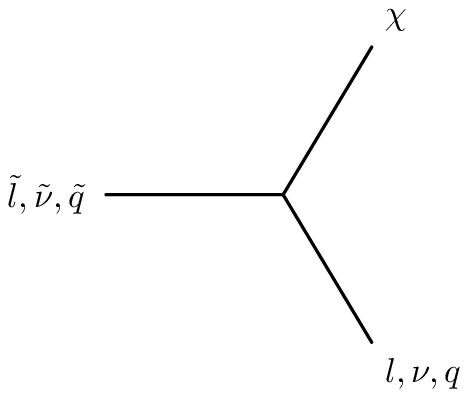}}
}
\end{center}
\caption[.]{ \small {\em Indirect} decays of
supersymmetric particles. The neutralino $\chi$ decays
directly via the
$\lambda$, $\lambda'$ and $\lambda''$ couplings as
 illustrated in Fig.~\ref{dec.examples}.}
\label{inddec.examples}
\end{figure}

The sparticle decays which proceed directly
to Standard Model particles are called {\it direct} 
decays (Fig.~\ref{dec.examples}). Decays in
which the sparticle first decays, conserving R-parity, to the lightest
neutralino are referred to as {\it indirect} decays (Fig.~\ref{inddec.examples}). 
Other cascade decays are possible
but not considered in the following.

The following assumptions are made throughout:
\begin{itemize}

\item{All three terms in Equation~(\ref{eqrpv}) are addressed, however
only one term at a time is considered to be nonzero for a specific set of indices 
($i$, $j$ and $k$). Unless otherwise stated the derived limits
correspond to the choice of indices for the coupling giving the least stringent
limit.}

\item{The lifetime of the sparticles can be neglected, \ie the mean
flight path is less than $1$~cm. This assumption restricts the sensitivity
of the search to R-parity violating couplings greater than  $10^{-4}$ for
gauginos and $10^{-7}$ for direct decays of sfermions, and constrains
the mass of the neutralino to be above 10 \gevcc~\cite{LLEpaper}.}

\item{Results are interpreted within the framework of the
MSSM with R-parity violation.
In addition to R-parity violating couplings, the parameters are 
 the gaugino mass terms ($M_i$), the sfermion masses  
($\ensuremath{m_{\tilde{\mathrm f}}}$), the ratio of the Higgs doublet 
vacuum expectation values ($\ensuremath{\mathrm{tan\beta}}$), the higgsino mass term 
($\ensuremath{\mathrm{\mu}}$) and the trilinear couplings ($\ensuremath{A_i}$).
Gaugino mass term unification at the 
GUT scale is assumed,
giving the condition $M_1={5\over3}M_2\tan^2\theta_W$} at the electroweak scale.

\item{For chargino and neutralino decays, only large values
 of sfermion masses are considered, with the consequence that the 
$direct$ decays of the lightest chargino and the next-to-lightest 
neutralino are suppressed. It also implies three-body kinematics 
for the lightest neutralino decay.  }

\end{itemize}

The searches presented in this paper cover all types of sparticle 
pair production; the case of single sneutrino production is addressed
in \cite{singlesneu}.
The results reported 
are based on all data 
collected by the ALEPH detector in the years 1998, 1999 and 2000. 
The new data collected in year 2000 at centre-of-mass energies up 
to $209~\gev$ is grouped in
two samples of $81.6~\invpb$ and $133.7~\invpb$
luminosity at $\langle \roots \rangle =204.9~
\gev$ and $206.5~\gev$,
respectively. Those two samples will be called $205~\gev$
and $207~\gev$ in the following.

This paper is organized as follows: after a brief description of the
ALEPH detector, the Monte Carlo generators
used for signal and background are listed in
Section~\ref{aleph.detector}. The selections used
for all topologies and the results obtained when those selections are applied 
on  data and Monte Carlo events are presented in Section \ref{selection}.
Section  \ref{interpretation} gives the interpretations, within the MSSM framework,
of the absence of any supersymmetric signal in the data.
A summary of the results is presented in Section~\ref{conclusions}.

\section{\label{aleph.detector}The ALEPH Detector and Monte Carlo generators}

The \ALEPH\ detector is described in detail in
Ref.~\cite{bib:detectorpaper}. An account of the performance of the
detector and a description of the standard analysis algorithms can be
found in Ref.~\cite{bib:performancepaper}. Here, only a brief
description of the detector components and the algorithms relevant for
this analysis is given.

The trajectories of charged particles are measured with a silicon
vertex detector, a cylindrical drift chamber, and a large time
projection chamber (TPC). The central detectors are immersed in a
1.5~T axial magnetic field provided by a superconducting solenoidal
coil.  The electromagnetic calorimeter (ECAL), placed between the TPC
and the coil, is a highly segmented sampling calorimeter which is used
to identify electrons and photons and to measure their energies. 
 The hadron calorimeter (HCAL) consists of the
iron return yoke of the magnet instrumented with streamer tubes. It
provides a measurement of hadronic energy and, together with the
external muon chambers, muon identification.
 The luminosity monitors extend the calorimetric coverage down to 34~mrad
from the beam axis. 
  The calorimetric and
tracking information are combined in an energy flow algorithm which
gives a measure of the total energy
with an uncertainty of $(0.6\sqrt{E}+0.6)$~GeV.

Electron identification is primarily based upon the matching between
the measured momentum of the charged track and the energy deposited in
the ECAL. Additional information from the shower profile in the ECAL
and the measured rate of specific ionisation energy loss in the TPC
are also used.  Muons are separated from hadrons by their
characteristic pattern in HCAL and the presence of associated hits in
the muon chambers.

The signal topologies were simulated using the {\tt SUSYGEN} Monte
Carlo program~\cite{susygen}, modified as described in 
Ref.~\cite{183paper}. The events were subsequently passed either through
a full simulation, or through a faster simplified simulation of the ALEPH
detector for interpolation purposes.

\pagebreak
Samples of all major backgrounds were generated and passed through the
full simulation. The {\tt PYTHIA} generator~\cite{pythia} was used to
produce $\qq$ events and four-fermion final states from $\PW\Pe\nu$,
$\PZ\PZ$ and $\PZ\Pe\Pe$. Pairs of W bosons were generated with {\tt
KORALW}~\cite{koralw}. Pair production of leptons was
simulated with {\tt BHWIDE}~\cite{bhwide} (electrons) and {\tt
KORALZ}~\cite{koralz} (muons and taus). The $\gamma\gamma \rightarrow
\mathrm{f\bar{f}}$ processes were generated with {\tt
PHOT02}~\cite{phot02}.

\section{\label{selection}Selections and results}

The selections were optimized to give the minimum expected $95\%$
C.L. excluded cross section in the absence of a signal for masses
close to the high end of the expected sensitivity. Selection
efficiencies were determined as a function of the SUSY particle masses
and the generation structure of the R-parity violating couplings
$\lambda_{ijk}$, $\lambda'_{ijk}$ and $\lambda^{''}_{ijk}$.
Two new selections have been added with respect to previous publications,
the others are unchanged from those used in
Refs.~\cite{LLEpaper,183paper,paper:202_rpv,LQDpaper,rpc183}, 
except for centre-of-mass energy rescaling.
The new selections : ``Many jets + Taus'' and ``Four jets + Taus''
address topologies with many jets and taus and are used in the search for indirect stau decays. 
The set of cuts is shown in Table~\ref{UDDcuts_stau}. 
Tau production is tagged by missing energy
and a low multiplicity jet. The corresponding event variable ($N^{min}_{jet-ch}$)
is the number of charged tracks in the lowest multiplicity jet when
forcing the event into four jets.

\begin{table}[hhh]
\caption{\small \label{UDDcuts_stau}
The list of cuts for the ``Four Jets + Taus''
and ``Many Jets + Taus'' selections, as used for the search for
indirect stau decays via a $\sudd$ operator. $N^{min}_{jet-ch}$ is
the number of charged tracks in the lowest multiplicity jet.
The other variables are defined in Ref.~\cite{183paper}} 
\begin{center}
\begin{tabular}{|c|c|} \hline
 Four Jets + Taus & Many Jets + Taus \\
\hline \hline
\multicolumn{2}{|c|}{$\nch\ > 8$}\\
\multicolumn{2}{|c|}{$|\pzmiss|/\pmiss<0.95$ , $0.5<E_\mathrm{vis}/\sqrt s<0.95$} \\
\multicolumn{2}{|c|}{$E^\mathrm{em}_\mathrm{jet} < 95\%E_\mathrm{jet}$ , $E_T>45~\gev$} \\
\multicolumn{2}{|c|}{$N^{min}_{jet-ch}<5$} \\
\hline
\hline
$0.6<T<0.92$               &  $0.6<T<0.97$                               \\
$y_4 >0.005$                &  $y_4 >0.005$                                \\
                            &  $y_6>0.0025$  \\
\hline
$|M_{12-34}|<10~\gevcc$    & \\
$|M_{12+34}-2M_{\stau}|< 5~\gevcc$ &  \\
\hline
\end{tabular}
\end{center}
\end{table}

\subsection {\label{lle} Selections for a dominant \boldmath{$\slle$} coupling}

Direct decays with a dominant $\slle$ coupling only involve leptons
and neutrinos 
in the final states. When indirect decays occur, additional leptons, neutrinos 
and jets are produced. 

The decay topologies consist either of purely leptonic 
final states --- as few as two acoplanar leptons
 in the simplest case (direct slepton decay) or as many
as six leptons plus four neutrinos in the most complicated case
(indirect chargino decay) --- or of  multi-jet and multi-lepton final states.

The various selections addressing the above topologies are summarized in
Table~\ref{topslle} together with the references of the papers in which 
detailed descriptions of the selection cuts can be found. The numbers of
selected
data candidates and the expected backgrounds are also given in this
table.

\begin{table}
\caption[.]{ \small   
The observed numbers of events in the year 2000 data sample and the corresponding
background expectations for the $\slle$ selections.
The selections are given together with the references to the papers in which 
they are described.}
\label{topslle}
\begin{center}
\renewcommand{\arraystretch}{1.1}
\begin{tabular}{|c|c|cc|cc|}
\hline
Selection & Ref. &
 \multicolumn{2}{|c|}{205~\gev} &
\multicolumn{2}{c|}{207~\gev} \\
 & & Data & Background & Data & Background  \\
\hline

Leptons and Hadrons& \cite{LLEpaper,paper:202_rpv} & 5 &    3.5 &    8 &    5.6 \\
\hline

6 Leptons + $\emiss$ & \cite{LLEpaper}&0 &    0.5 &    0 &    0.8 \\
\hline

4 Leptons + $\emiss$ &\cite{LLEpaper} & 1 &    2.3 &    4 &    3.1 \\
\hline

 $\ell\ell\ell\ell$ &\cite{LLEpaper} & 1 &    3.8 &    3 &    3.5 \\
\hline

$\ell\ell\tau\tau$& \cite{LLEpaper}& 0 &    1.1 &    1 &    1.4 \\
\hline

$\tau\tau\tau\tau$ & \cite{LLEpaper}& 1 &    2.4 &    5 &    2.6 \\
\hline

Acoplanar Leptons & \cite{LLEpaper,183paper}& 71 &   89 &  139 &  138 \\
\hline
\end{tabular}
\renewcommand{\arraystretch}{1.}
\end{center}
\end{table}

%***********************************************************************
\subsection {\label{lqd} Selections for a dominant \boldmath{$\slqd$} Coupling}

 For a dominant $\slqd$ operator the event topologies are mainly
characterized by large hadronic activity, possibly with some leptons
and some missing energy. In the simplest case the topology consists of
four jets, and in the more complicated scenarios, of
several jets or leptons, with or without missing mass.
The various selections are listed in Table~\ref{topslqd}
together with the corresponding numbers of observed data candidates and
expected background events.

\subsection {\label{udd} Selections for a dominant \boldmath{\sudd}  coupling}

For a dominant $\sudd$ operator the final states are characterized 
by topologies having many hadronic jets, possibly associated with 
leptons or taus and missing energy.

These selections rely mainly on two
characteristics of the signal events: the reconstructed mass of pair
produced sparticles
and the presence of many jets in the final state. In Table~\ref{topsUDD}
 the list of
all the selections is given, together with the numbers of data candidates and 
expected background events.

\begin{table}[hhh]
\caption[.]{ \small 
The observed numbers of events in the year 2000 data sample and the corresponding
background expectations for the $\slqd$ selections.
The selections are given together with the references to the papers in which 
they are described.}
\label{topslqd}
\setlength{\tabcolsep}{0.12cm}
\begin{center}
\renewcommand{\arraystretch}{1.1}
\begin{tabular}{|c|c|cc|cc|}
\hline
Selection & Ref.&
 \multicolumn{2}{|c|}{205~\gev} &
\multicolumn{2}{c|}{207~\gev} \\
 & & Data & Background & Data & Background  \\
\hline
MultiJets + Leptons & \cite{paper:202_rpv,LQDpaper} &5 &    5.2 &   14 &    8.6 \\
\hline

Jets-HM & \cite{paper:202_rpv} &  3 &    2.2 &    7 &    3.3 \\
\hline

4 Jets + 2$\tau$ & \cite{paper:202_rpv,LQDpaper} & 9 &    5.0 &    6 &    8.0 \\
\hline

Four-Jets & \cite{paper:202_rpv,LQDpaper} &341 &  348 &  541 &  561 \\
\hline

2 Jets + 2$\tau$ & \cite{paper:202_rpv,LQDpaper}& 7 &    4.9 &    7 &    7.9 \\
\hline

AJ-H & \cite{rpc183} & 12 &   10.7 &   19 &   18.5 \\
\hline

4JH & \cite{rpc183} &4 &    3.7 &    4 &    5.9 \\
\hline

5 Jets + 1 Iso. $\ell$ &\cite{paper:202_rpv} &1 &    2.1 &    2 &    3.7 \\
\hline

4 Jets + 2 Iso. $\ell$ &\cite{paper:202_rpv}& 0 &    1.3 &    1 &    2.0 \\
\hline
\end{tabular}
\renewcommand{\arraystretch}{1.}
\end{center}
\end{table}

\begin{table}[hhh]
\caption[.]{ \small 
The observed numbers of events in the year 2000 data sample and the corresponding
 background expectations for the $\sudd$ selections.
The selections are given together with the references to the papers in which 
they are described.}

\label{topsUDD}
\begin{center}
\begin{tabular}{|c|c|cc|cc|}
\hline
Selection & Ref. &
 \multicolumn{2}{|c|}{205~\gev} &
\multicolumn{2}{c|}{207~\gev} \\
 & & Data & Background & Data & Background  \\

\hline
Four Jets  Broad &\cite{183paper}& 53 & 51.8 & 72 & 84.1 \\
\hline
Many Jets &\cite{183paper}& 6 & 3.8 & 6 & 6.2 \\
%% Many Jets     &  7 &   5.27 &    8 &    8.32  \\ 
\hline
Many Jets + Leptons &\cite{183paper}& 6 & 7.6 & 14 & 12 \\
\hline
Four Jets + 2 Leptons   &\cite{183paper}& 2 & 2.1 & 4 & 3.6 \\
\hline
Many Jets + 2 Leptons  &\cite{183paper}& 3 & 2.9 & 1 & 4.9 \\
\hline
Four Jets+$\emiss$  &\cite{183paper,paper:202_rpv}& 32 & 33.3 & 48 & 51.4 \\
\hline
Many Jets+$\emiss$ &\cite{183paper,paper:202_rpv}& 30 & 33.3 & 48 & 51.4 \\
\hline
Four Jets + Taus   &  & 78 & 76.1 & 144 & 125.5 \\
\hline
Many Jets + Taus  & & 8 & 9.7 & 17 & 15.0 \\
\hline

\end{tabular}
\end{center}
\end{table}

\section{\label{interpretation}Interpretation within the MSSM framework}

For all selections,
the number of candidate events observed in the data is in agreement with the 
Standard Model background expectations.
The results of the selections have been used to set limits on the MSSM parameter
space.

The cross-section limits were evaluated at 208~\gev.
Data taken at lower centre-of-mass energies also contribute
to the limits with a reduced weight. The weight was calculated from
the expected evolution of the cross section with $\roots$.

The systematic uncertainties on the selection efficiencies are of the
order of 4--5$\%$ and are dominated by the 
statistics of the simulated signal samples,
with small additional contributions
from lepton identification and energy flow simulation. They were
taken into account by reducing the selection efficiencies by 
the estimated systematic uncertainty.

When setting the limits, background subtraction was performed for two-
and four-fermion final states according to the prescription described in
Ref.~\cite{PDG}.
The systematic uncertainty on the expected Standard Model background has been
evaluated by detailed comparison of the simulation with the data on 
control samples obtained with relaxed cuts.
The subtracted background has been reduced by the systematic uncertainty
derived from these comparisons (typically a few percent depending on the analysis).
For the $\PW\Pe\nu$ and $\PZ\Pe\Pe$ processes the subtracted background has
 been further reduced by $20\%$ due to the poor knowledge of the production
cross section in the kinematic region selected by this analysis.
No background is subtracted for the 
$\gamma\gamma \rightarrow \mathrm{f\bar{f}}$ process.

The absolute lower limit on the mass of the lightest neutralino of
$23~\gevcc$ obtained in Ref.~\cite{LLEpaper} for a dominant $\slle$ coupling, 
which is valid for any
choice of $\mu$, $M_2$, $m_0$ (the unified sfermion mass term at the GUT scale) 
and generational indices ($i$, $j$ and
$k$), is used to restrict the range of neutralino mass considered for
the indirect decays of the $\slle$ searches.

The limits on sfermion masses derived from searches and indirect
constraints obtained at LEP1 are discussed in Ref.~\cite{183paper}. 
They range from 40 to 45~\gevcc\ and are indicated on the exclusion plots.

\subsection {Charginos and neutralinos decaying via \boldmath $\slle$}
\label{challe}

The results are interpreted assuming large
scalar masses ($m_0=500~\gevcc$).
Depending on the masses of the gauginos and on the lepton flavour
composition in the decay, the indirect decays of charginos to
neutralinos populate
different regions in track multiplicity, visible mass and leptonic
energy. The ``Leptons and Hadrons'' selection, optimized for each 
possible topology, is used.

\begin{figure}
\begin{center}
\resizebox{\textwidth}{!}{
  \includegraphics{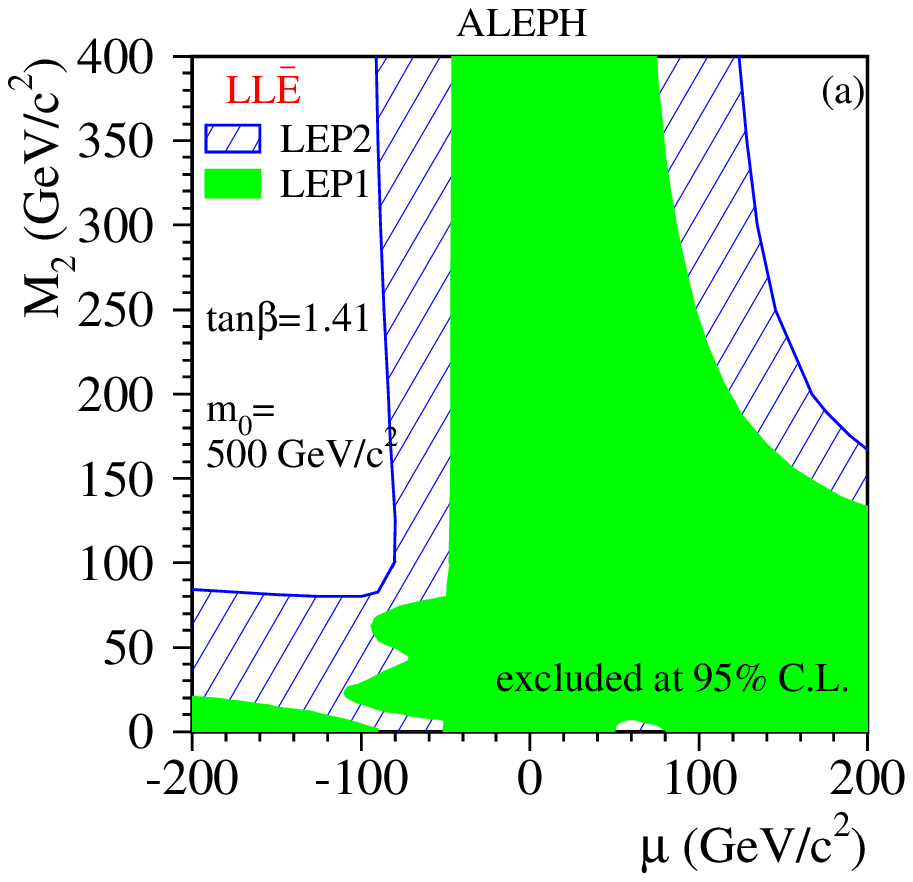}
  \includegraphics{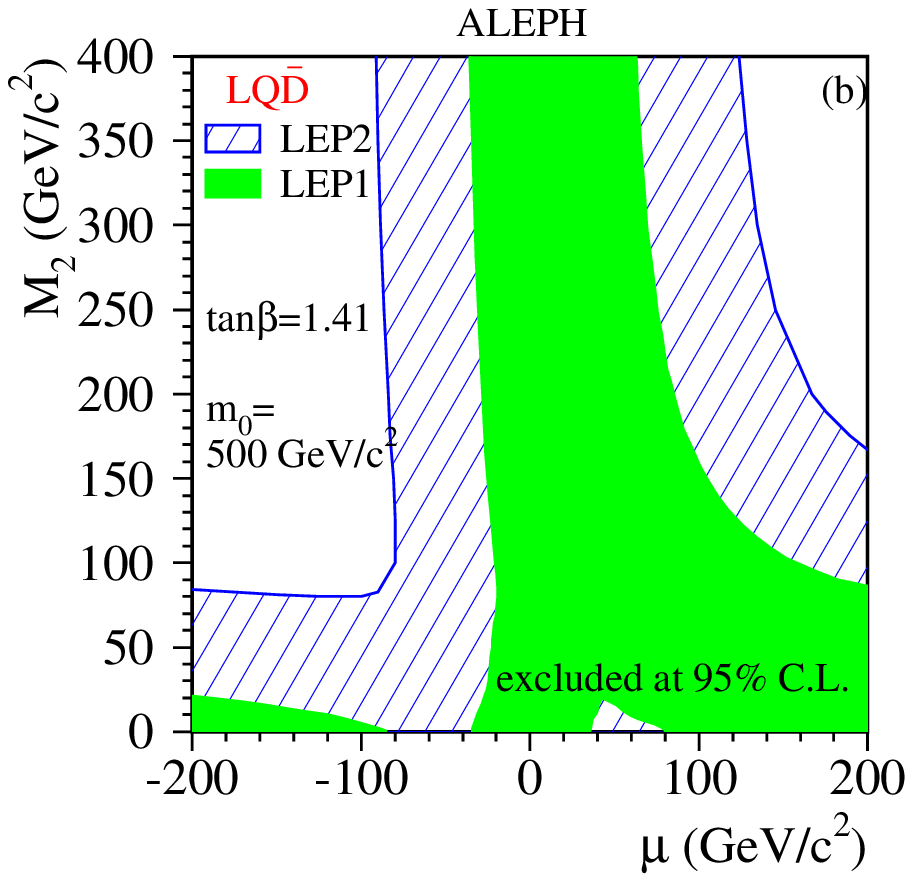}
}
\resizebox{0.5\textwidth}{!}{
  \includegraphics{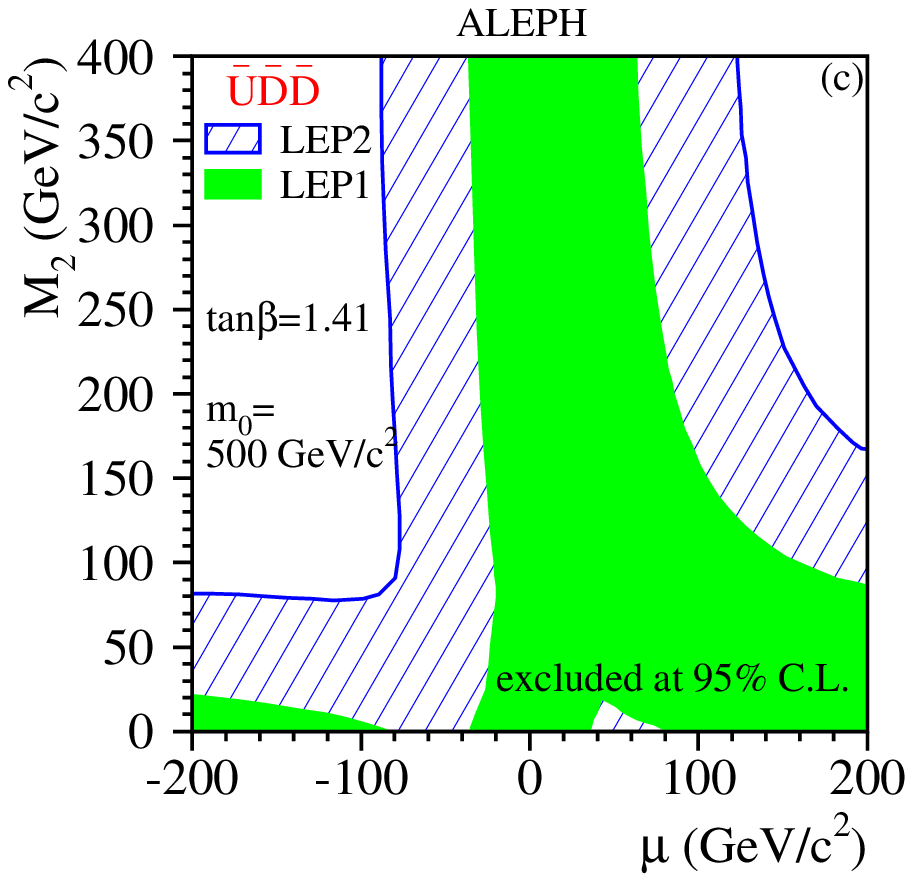}
}   
\caption[.]{  \label{mum2lims}{\small Regions in the $(\mu,M_2)$
plane excluded at $95\%$ C.L. at $\tanb~=~1.41$ and \\ $m_0~=~500~\gevcc$ for
the three operators (a) $\lambda$, (b) $\lambda'$ and (c) $\lambda''$. }}
  \end{center}
\end{figure}

In the
framework of the MSSM, 95$\%$ C.L. exclusion limits are derived in the
($\mu$,$M_2$) plane as shown in Fig.~\ref{mum2lims}a for $\tanb = 1.41$.
The corresponding lower limit on the
mass of the lightest chargino is  103~\gevcc.

The searches for the lightest and second lightest
neutralino do not extend the excluded region in the
($\mu$,$M_2$) plane beyond that achieved with the chargino 
search alone.

\subsection{Squarks decaying via \boldmath{$\slle$}}
\begin{figure}
\begin{center}
\resizebox{\textwidth}{!}{
  \includegraphics{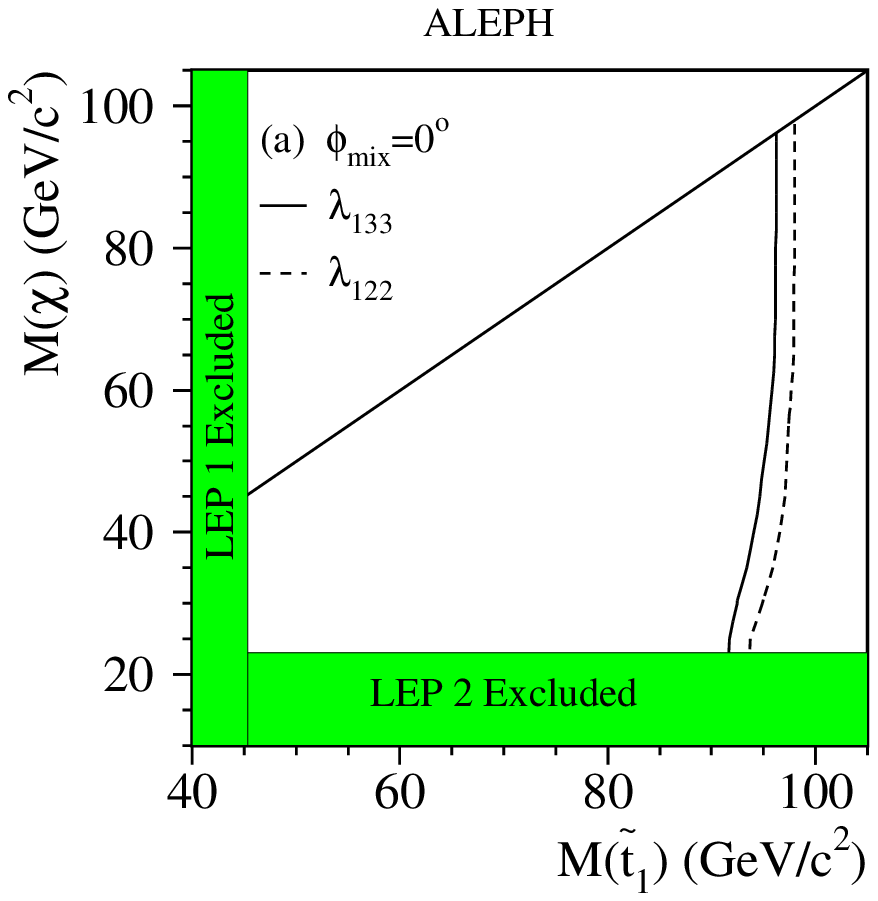}
  \includegraphics{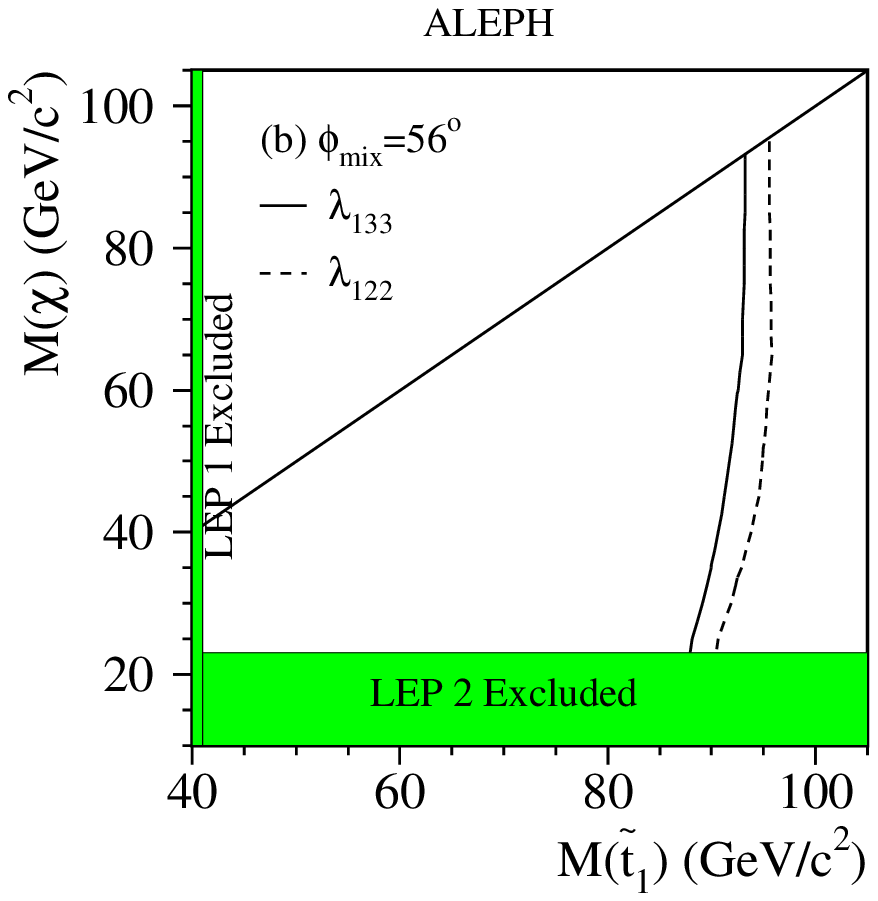}
}
\resizebox{\textwidth}{!}{
  \includegraphics{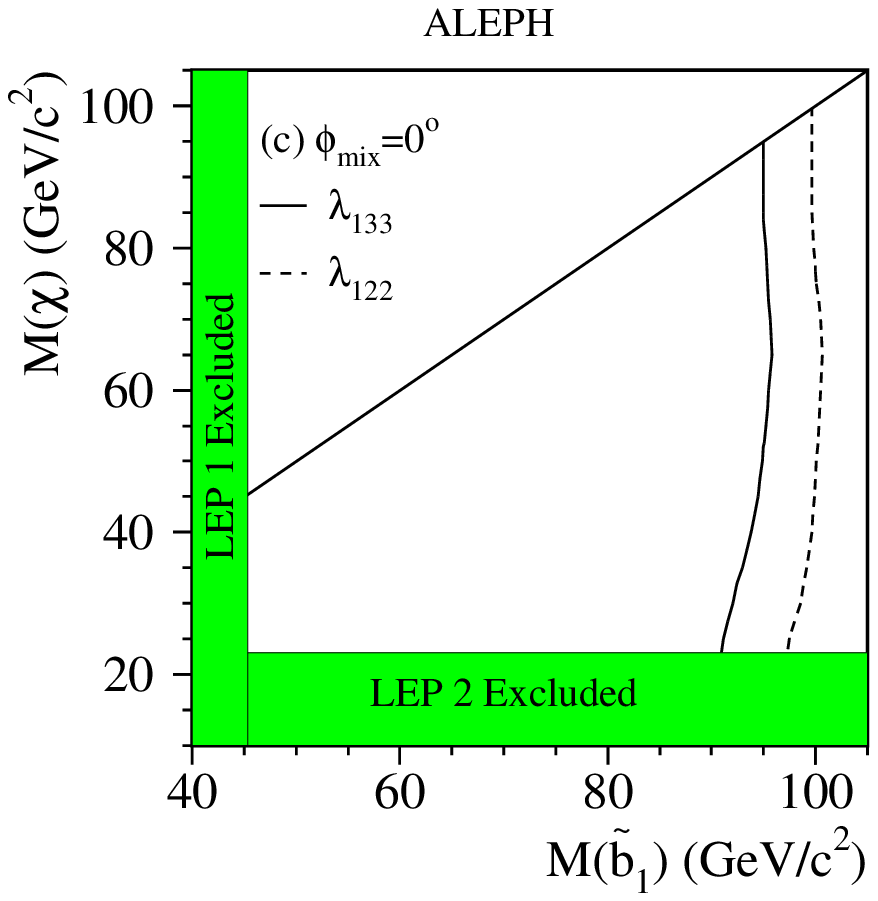}
  \includegraphics{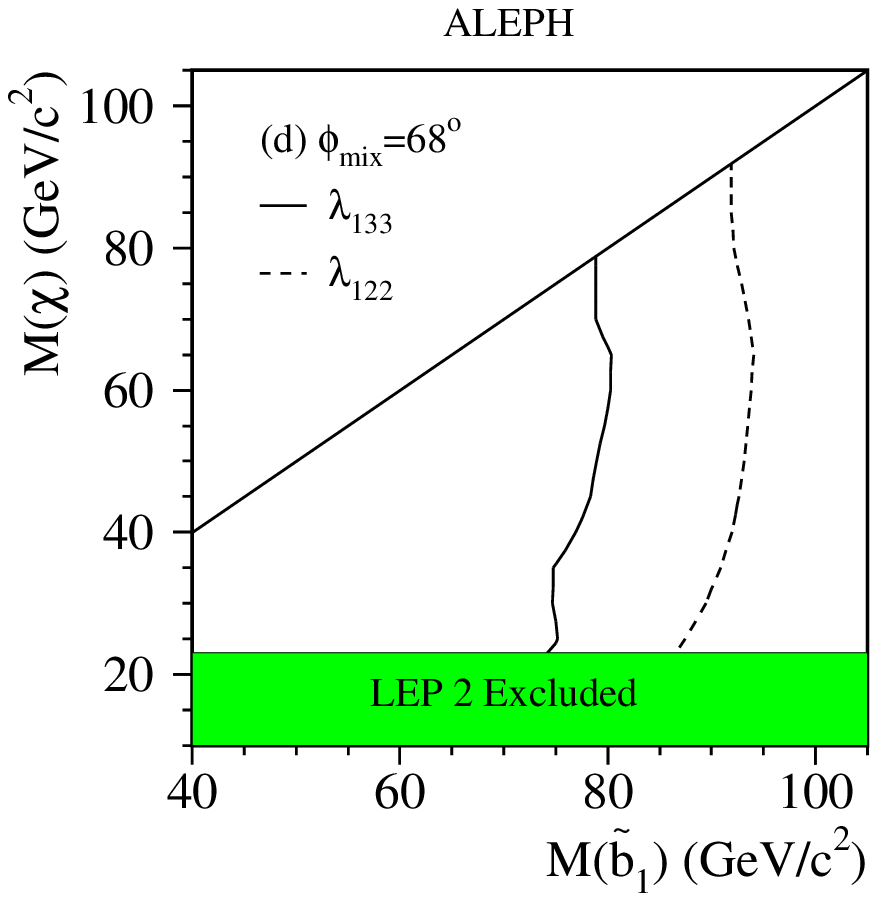}
}
\caption[.]{  \label{lle_stops}{\small The $95\%$ C.L. limits in
(a),(b) the ($m_{\chi}$, $m_{\stone}$) plane and (c),(d) the ($m_{\chi}$, $m_{\sbone}$) 
plane for
indirect decays via the $\slle$ couplings 
$\lambda_{122}$ and $\lambda_{133}$, 
for no mixing ($\phimix=0^\circ$) and for $\phimix=56^\circ$ and $68^\circ$,
corresponding to vanishing coupling to the Z,
for stops and sbottoms, respectively. The LEP 2 exclusion corresponds 
to the absolute limit on $m_{\chi}$.}}
\end{center}
\end{figure}

 Squarks cannot decay directly with an $\slle$ coupling
but they may decay indirectly via the lightest neutralino.
Because the resulting topology
is close to that arising from the indirect chargino decay,  
the ``Leptons and Hadrons'' selection is used.  The
$95\%$ C.L. squark mass limits are presented as functions of
$m_{\chi}$ in Fig.~\ref{lle_stops} for the case of $\stone$ and
$\sbone$ squarks. 
The results are displayed for left-handed squarks and for the values of
the mixing angle for which the coupling to the Z vanishes.
In the case of purely left-handed squarks, the following limits
can be derived: $m_{\stL}>91~\gevcc$ and $m_{\sbL}>90~\gevcc$ for any
$\lam_{ijk}$.

\subsection{Sleptons decaying via \boldmath{$\slle$}}

\begin{figure}
\begin{center}
\resizebox{\textwidth}{!}{
  \includegraphics{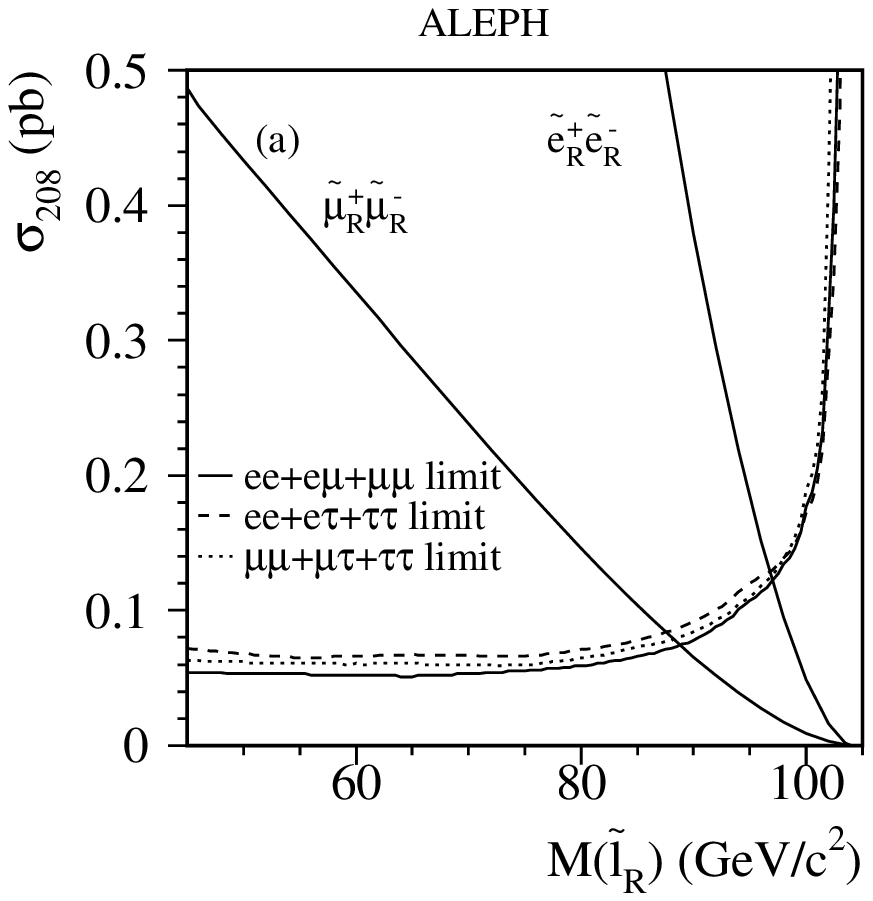}
  \includegraphics{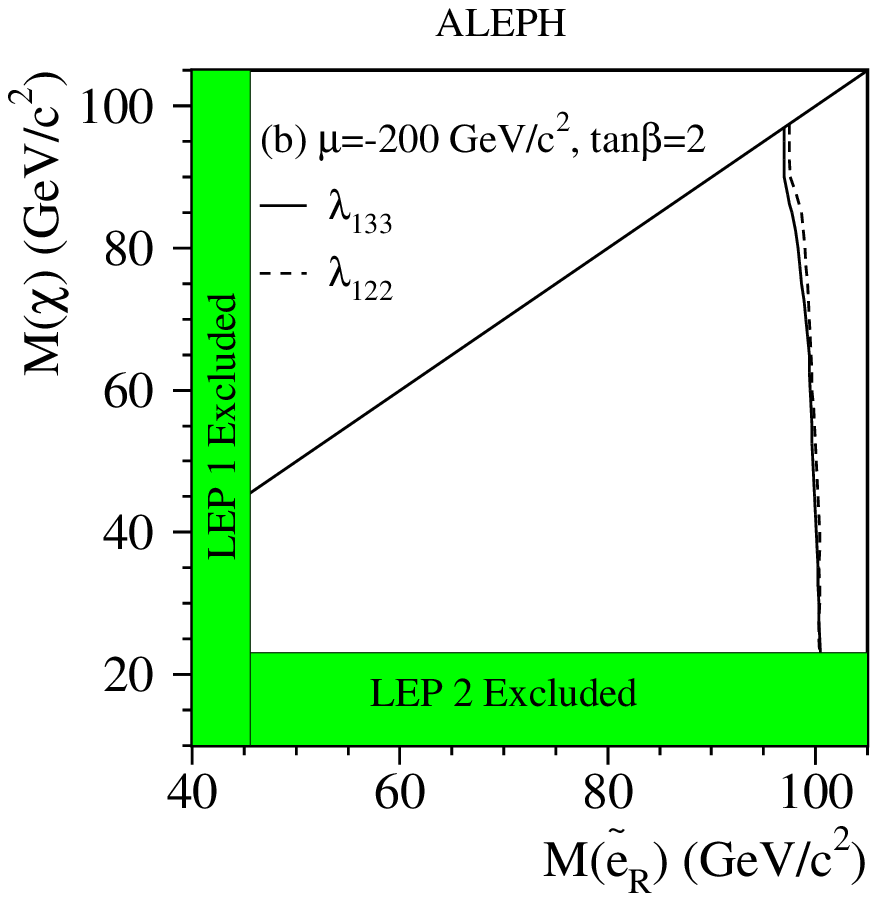}
}
\resizebox{\textwidth}{!}{
  \includegraphics{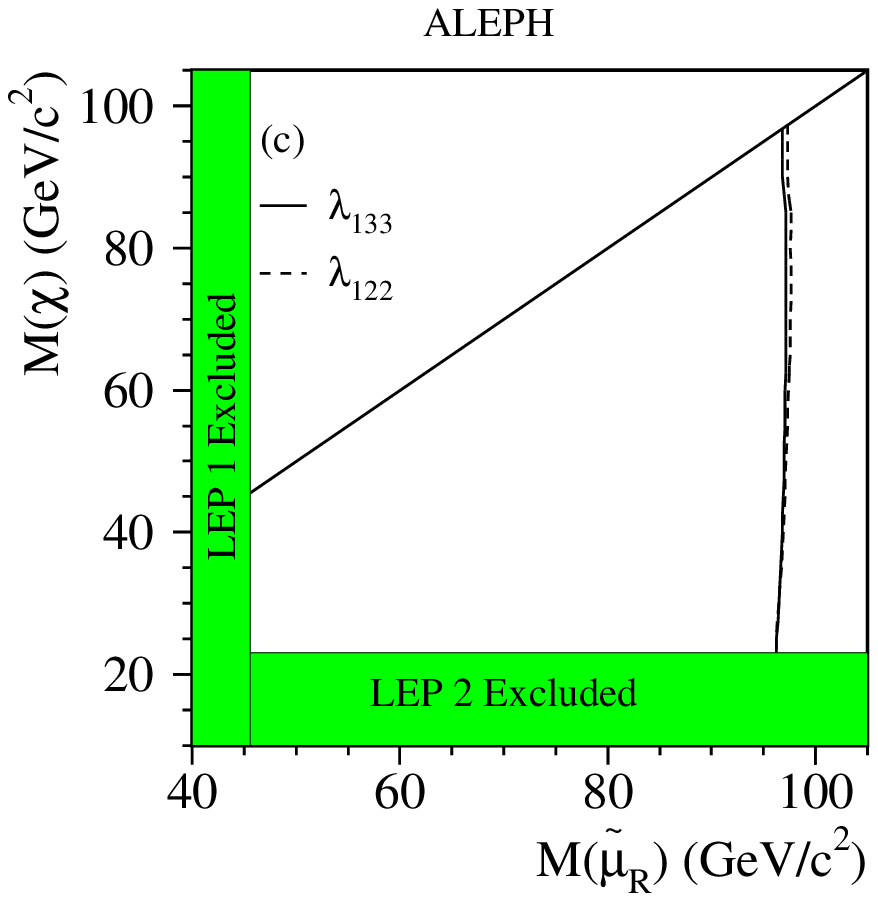}
  \includegraphics{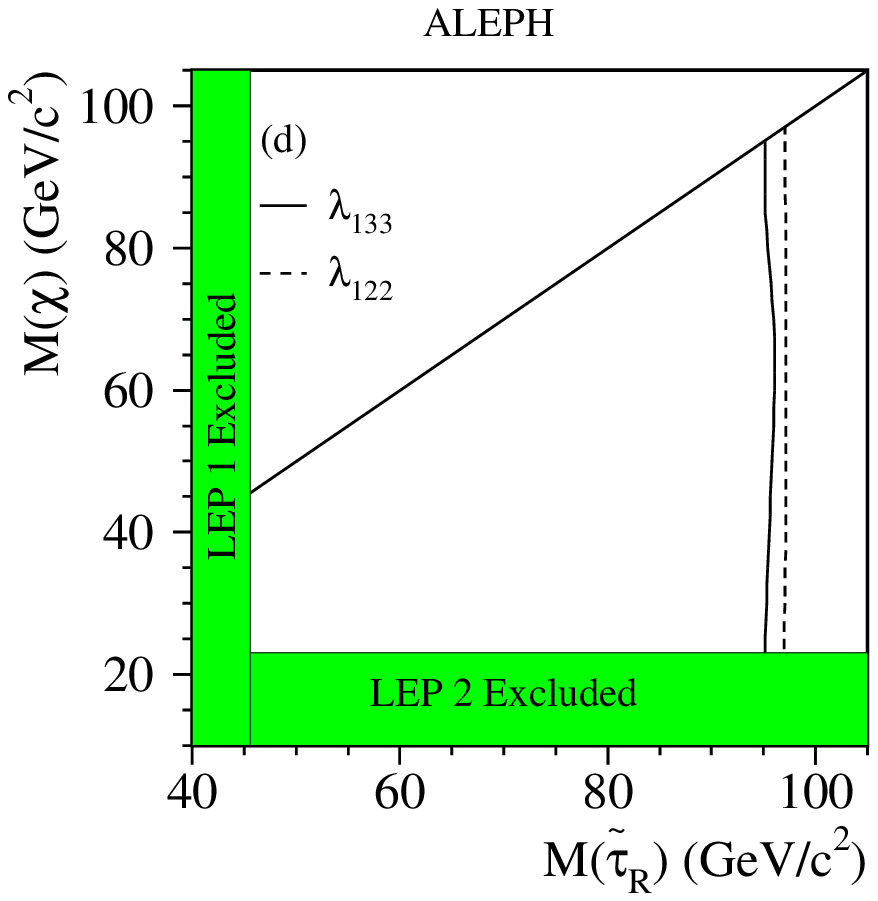}
}
\caption[.]{  \label{lle_sleptons}{(a) \small The $95\%$ C.L. 
cross-section upper limits for sleptons decaying directly via a 
dominant $\slle$
operator. The MSSM cross sections for pair production of right-handed
selectrons and smuons are superimposed. The
$95\%$ C.L. limits in the ($m_{\chi}$, $m_{\slR}$) plane for indirect
decays of selectrons (b), smuons (c) and staus (d). The two choices of
$\lambda_{122}$ and $\lambda_{133}$ correspond to the most and least stringent
exclusions, respectively. The selectron cross section is
evaluated at $\mu=-200~\gevcc$ and $\tanb=2$.}}
\end{center}
\end{figure}

A slepton can decay directly via the $\slle$
coupling to a lepton and anti-neutrino, hence the ``Acoplanar Leptons''
selection is used. For a given choice of {$\slle$} coupling, 
the decay of a right-handed slepton
produces two different final states, 
${\tilde{\ell}_R}^k \ra
\ell^i{\bar{\nu}_{\ell^j}}$ or ${\bar{\nu}_{\ell^i}}\ell^j$, with equal branching ratios.
  Excluded cross
sections are shown in Fig.~\ref{lle_sleptons}a for the different
mixtures of acoplanar lepton states. The MSSM production cross sections
for right-handed smuon pairs, and for selectron pairs at $\mu =
-200~\gevcc$ and $\tanb=2$, are superimposed. The cross section limits
translate into lower bounds of
$m_{\smuR,\stauR}>87~\gevcc$ and $m_{\seR}>96~\gevcc$ ($\mu =
-200~\gevcc$, $\tanb=2$).

  Indirect decays of sleptons are selected using the ``Six Leptons $+
\emiss$'' selection. Limits corresponding to this case are shown in
Figs.~\ref{lle_sleptons}b, \ref{lle_sleptons}c and \ref{lle_sleptons}d.  Using the bound of
$m_{\chi}>23~\gevcc$ these limits can be interpreted as the mass
limits $m_{\seR}>96~\gevcc$ ($\mu=-200~\gevcc$, $\tanb=2$),
$m_{\smuR}>96~\gevcc$ and $m_{\stauR}>95~\gevcc$.

\subsection{Sneutrinos decaying via \boldmath{$\slle$}}

\begin{figure}
\begin{center}
\resizebox{\textwidth}{!}{
  \includegraphics{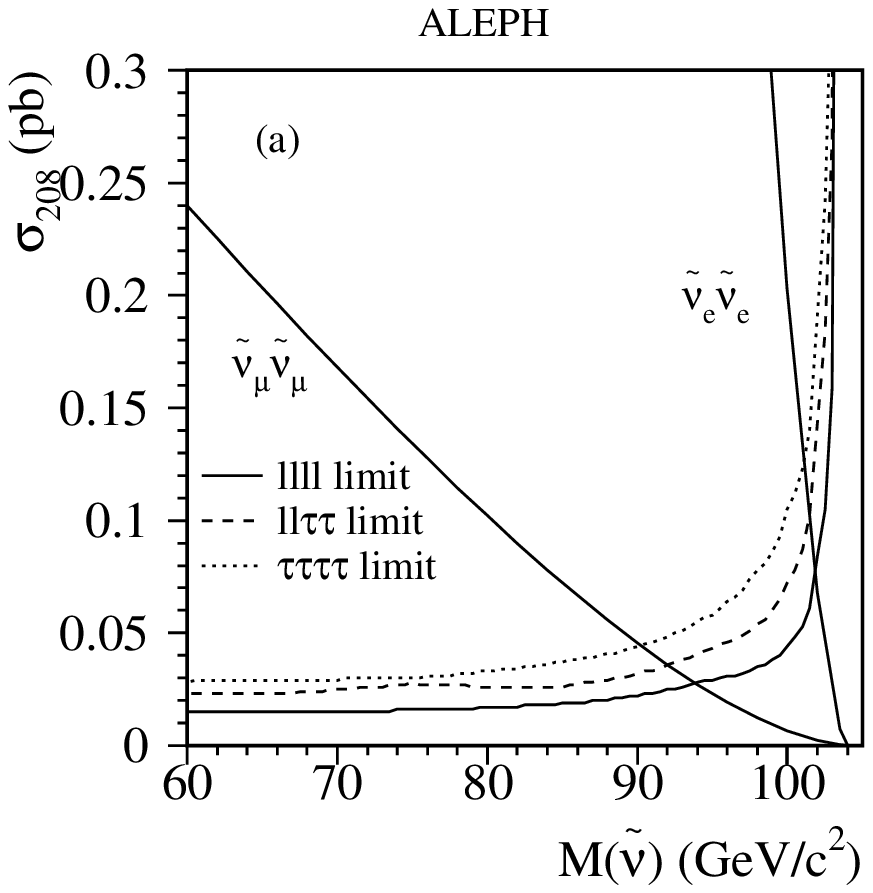}
  \includegraphics{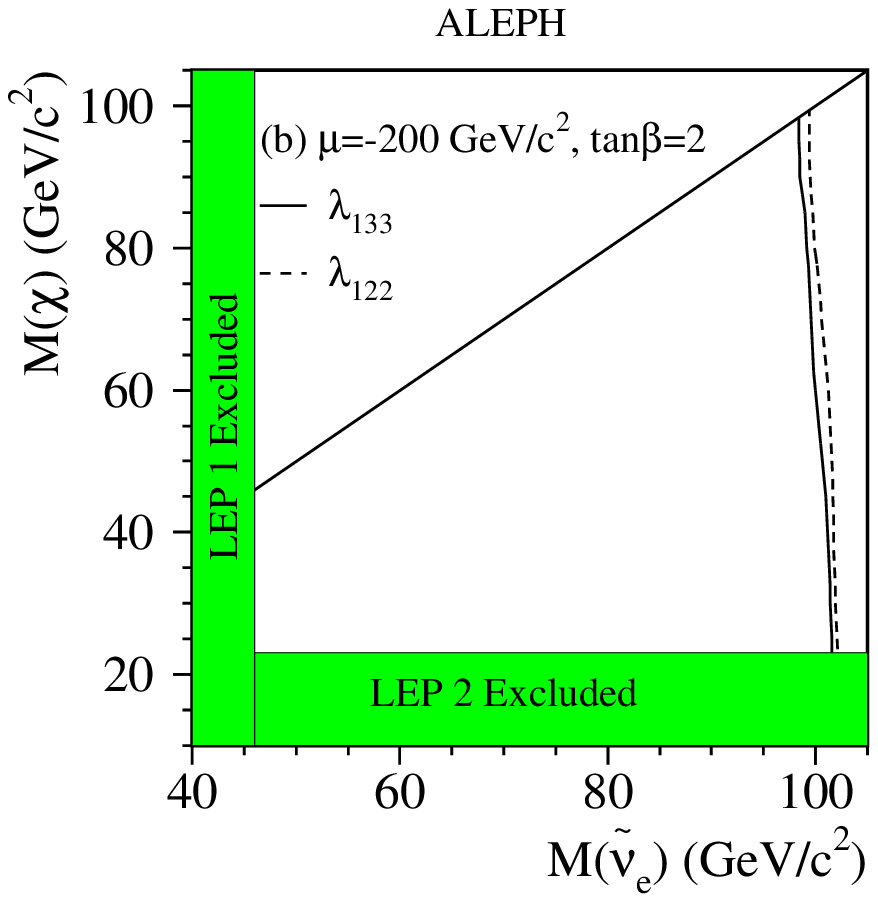}
}
\resizebox{0.5\textwidth}{!}{
  \includegraphics{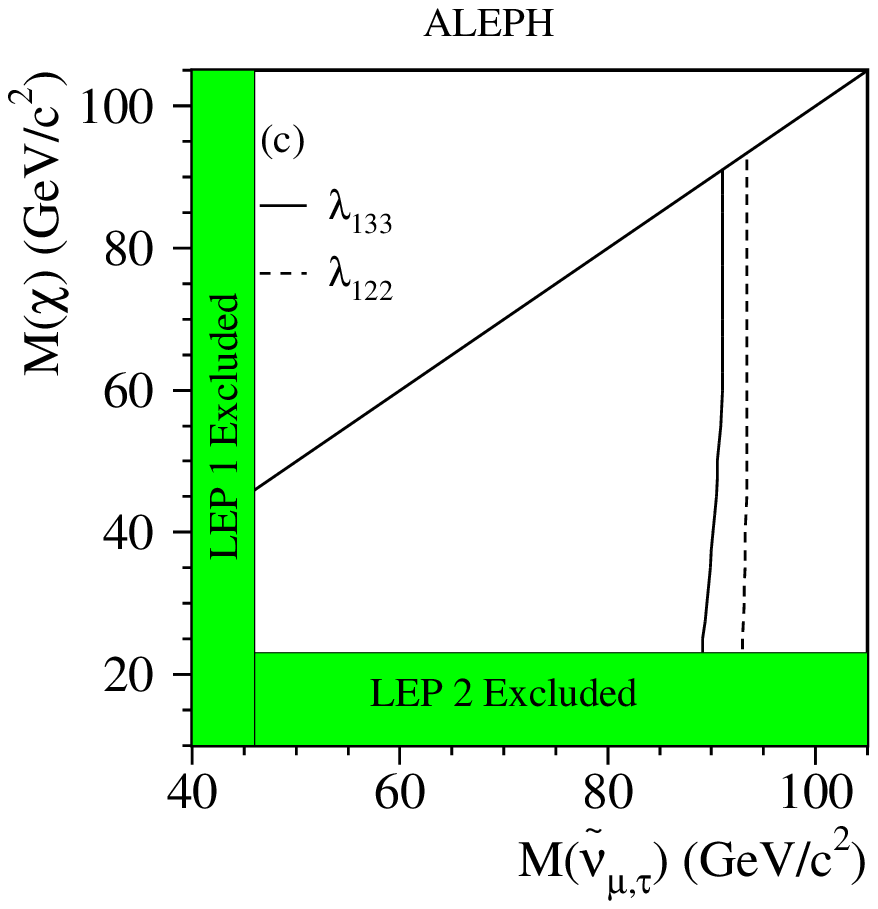}
}
\caption[.]{  \label{lle_sneus}{(a) \small The $95\%$ C.L. 
cross-section upper limits for sneutrinos decaying directly via a 
dominant $\slle$
operator. The three curves correspond to different possible final
states, with $\ell = {\mathrm e}$ or $\mu$, due to the specific choice of sneutrino
flavour and $\lambda_{ijk}$. The MSSM cross section for pair
production of muon and electron sneutrinos are superimposed; the tau
sneutrinos have the same cross section as the muon type. 
The $95\%$ C.L. limits in the ($m_{\chi}$, $m_{\snu}$) plane for
$\snu_\mathrm{e}$ (b) and for both $\snu_{\mu}$ and $\snu_{\tau}$ (c) 
indirect decays. The two choices of $\lambda_{122}$ and
$\lambda_{133}$ correspond to the most and least stringent exclusions,
respectively. The electron sneutrino cross section is evaluated at
$\mu=-200~\gevcc$ and $\tanb=2$.}}
\end{center}
\end{figure}

  Sneutrinos can decay directly via {$\slle$}
into pairs of charged leptons. For pair produced sneutrinos, 
the final states, depending on the generation indices, are eeee, ee$\mu\mu$,
ee$\tau\tau$, $\mu \mu \mu \mu$, $\mu \mu \tau \tau$ and $\tau \tau
\tau \tau$, and can be selected with the  ``Four Lepton'' selection.
 The exclusion limits on the sneutrino pair production
cross section are shown in Fig.~\ref{lle_sneus}a.  These limits
translate into a lower bound on the electron sneutrino mass of
$m_{\snue}>100~\gevcc$ ($\mu=-200~\gevcc$, $\tanb=2$) and the muon
sneutrino mass of $m_{\snumu}>90~\gevcc$.

  Indirect decays of sneutrinos are selected using the ``Four Leptons
$+ \emiss$'' selection. The limits in the ($m_{\chi}$, $m_{\snu}$)
plane corresponding to this case are shown in Figs.~\ref{lle_sneus}b
and \ref{lle_sneus}c.  Using the bound $m_{\chi}>23~\gevcc$
this limit can be interpreted as $m_{\snu_{\mu,\tau}} > 89~\gevcc$ and
$m_{\snue} > 98~\gevcc$, where the cross
section for the electron sneutrino is evaluated at $\mu=-200~\gevcc$
and $\tanb=2$.

\subsection{Charginos and neutralinos decaying via \boldmath{\slqd}}

The results are interpreted assuming large
scalar masses ($m_0=500~\gevcc$).
For the various
topologies produced in the indirect decays of the chargino pairs via
an $\slqd$ coupling, the ``MultiJets + Leptons'' selection is used.

 In the framework
of the MSSM, $95\%$ C.L. exclusion limits are derived in the
($\mu$,$M_2$) plane as shown in Fig.~\ref{mum2lims}b. 
The corresponding lower limit on the
mass of the lightest chargino is 103~\gevcc.

The searches for the lightest and second lightest
neutralino do not extend the excluded region in the
($\mu$,$M_2$) plane beyond that achieved with the chargino 
search alone.

\subsection{Squarks decaying via \boldmath{\slqd} }

\begin{figure}
\begin{center}
\resizebox{\textwidth}{!}{
  \includegraphics{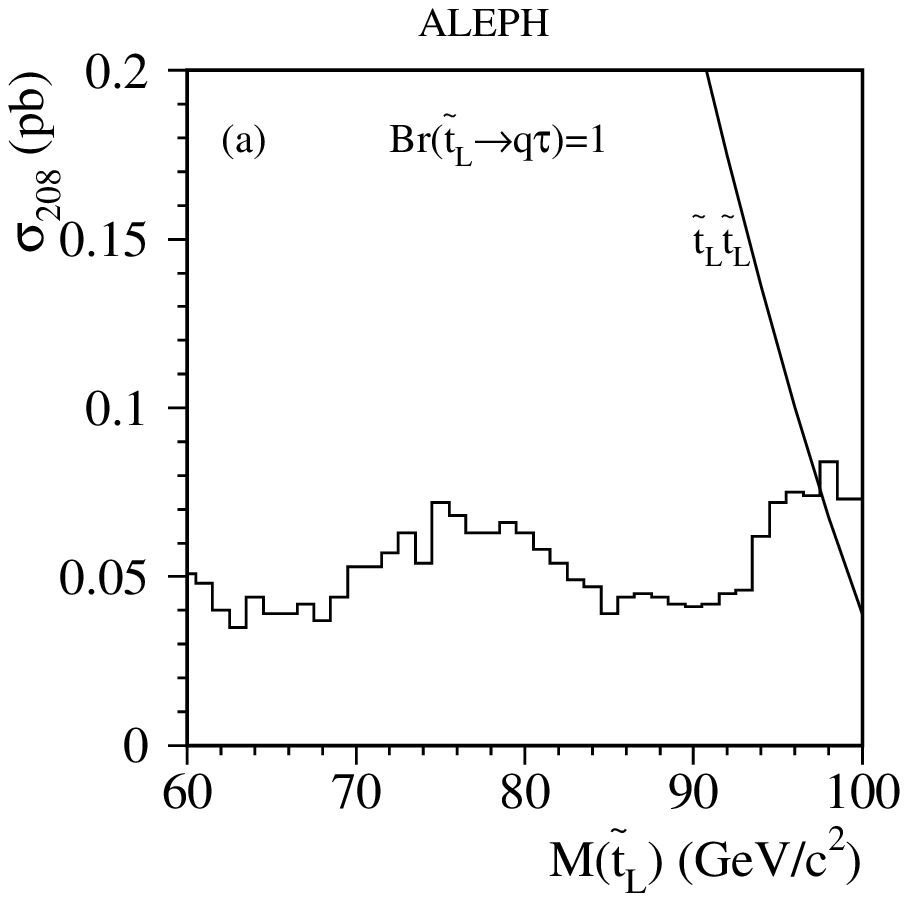}
  \includegraphics{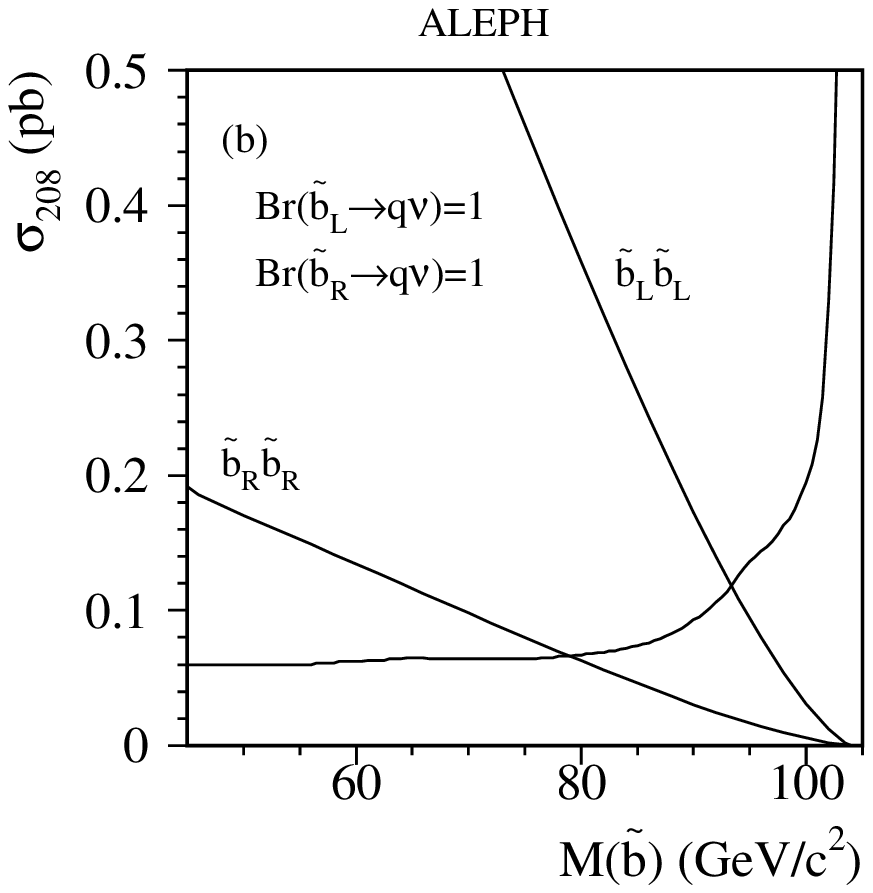}
}
\resizebox{0.5\textwidth}{!}{  
  \includegraphics{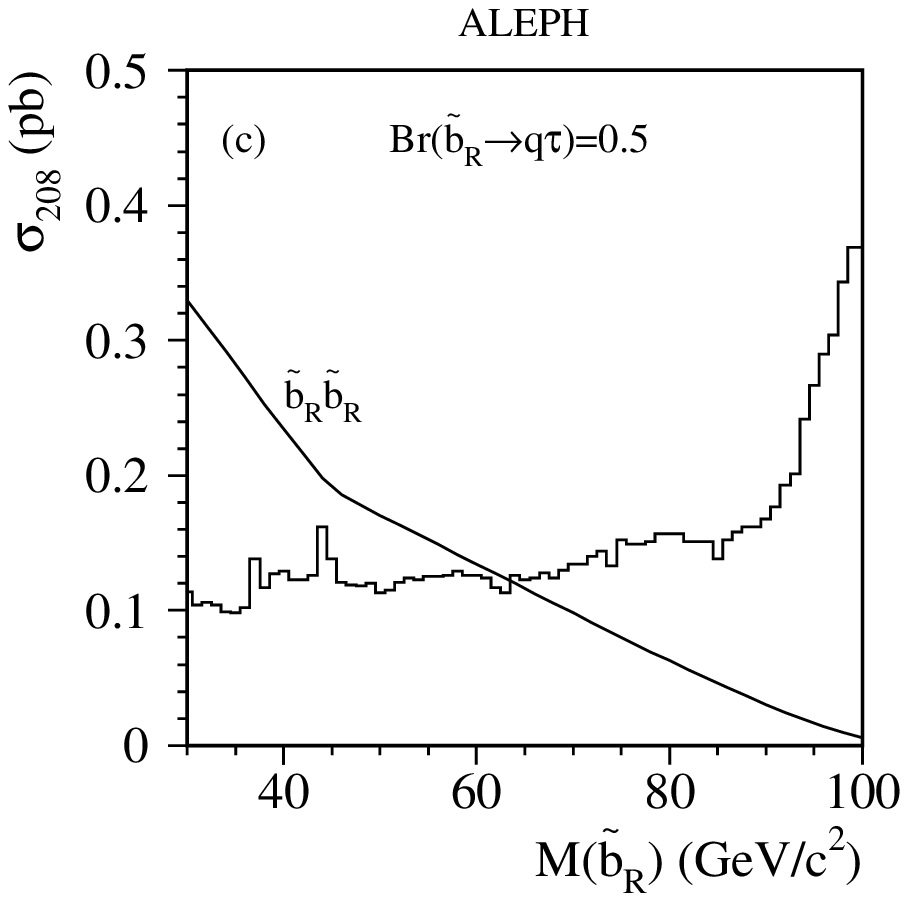}
}
\caption[.]{  \label{sq.direct}{\small 
The $95\%$ C.L. cross section upper limits for the production of squarks
decaying directly via a dominant $\slqd$ operator: the limits are shown for $\stL$
$(\lambda'_{33k})$, $\sbL$ $(\lambda'_{i3k})$ or $\sbR$
$(\lambda'_{i33})$, and $\sbR$~$(\lambda'_{3j3})$ in (a), (b) and (c), respectively.
The expected MSSM cross sections are superimposed.}}
\end{center}
\end{figure}

A squark can decay directly via {\slqd} to a quark and either a lepton or a
neutrino, leading to topologies with acoplanar jets and up to two
leptons. Couplings with electrons or muons in the final state are not
considered as existing limits from the Tevatron~\cite{TeVLQ} exclude
the possibility of seeing such a signal at LEP.  To select $\sq \ra \q
\tau$ and $\sq \ra \q\nu$, the ``2J+2$\tau$'' and the ``AJ-H''
selections are used.   Examples of limits for
squark production are shown in Fig.~\ref{sq.direct}. In particular,
for a dominant $\lambda'_{33k}$ coupling, which implies Br$(\stL
\rightarrow \mathrm{q} \tau)=100\%$, a lower limit of $m_{\stL}>97~\gevcc$ is
obtained.

 Indirect decays of squarks via the $\slqd$ operator lead to
six jets and up to two charged leptons. 
The selections used are ``Jets-HM'',``4 Jets + 2$\tau$'',
 ``5 Jets + 1 Iso.~$\ell$'' and ``Multi-jets plus Leptons''.

\begin{figure}
\begin{center}
\resizebox{\textwidth}{!}{
  \includegraphics{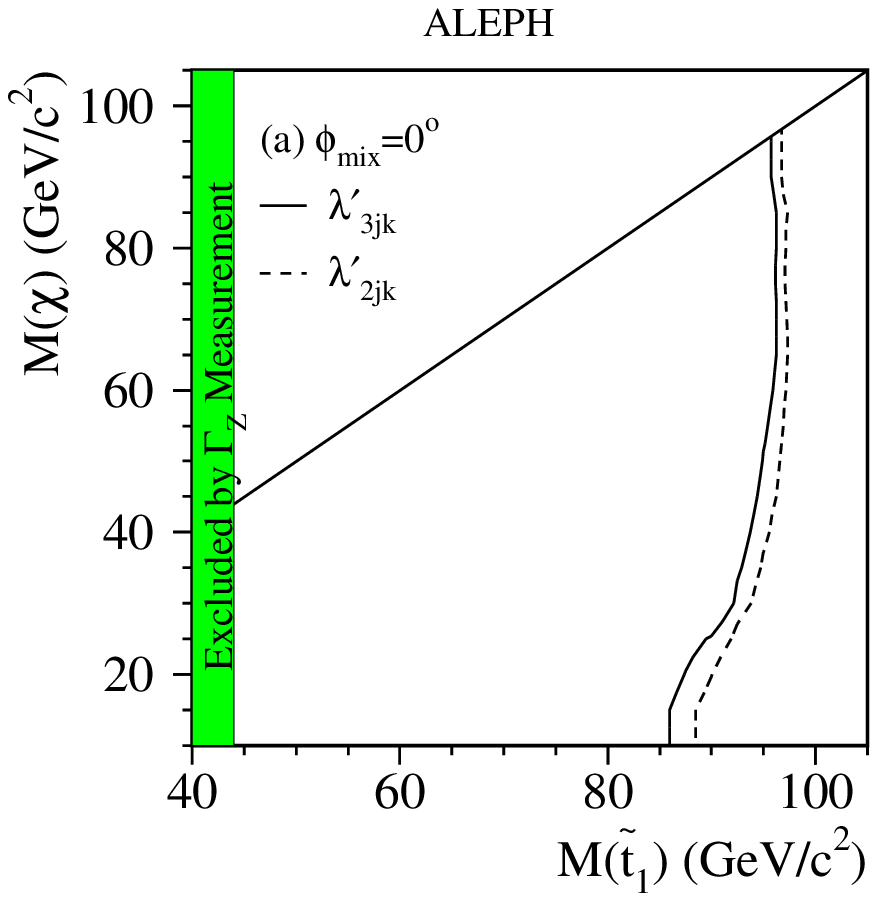}
  \includegraphics{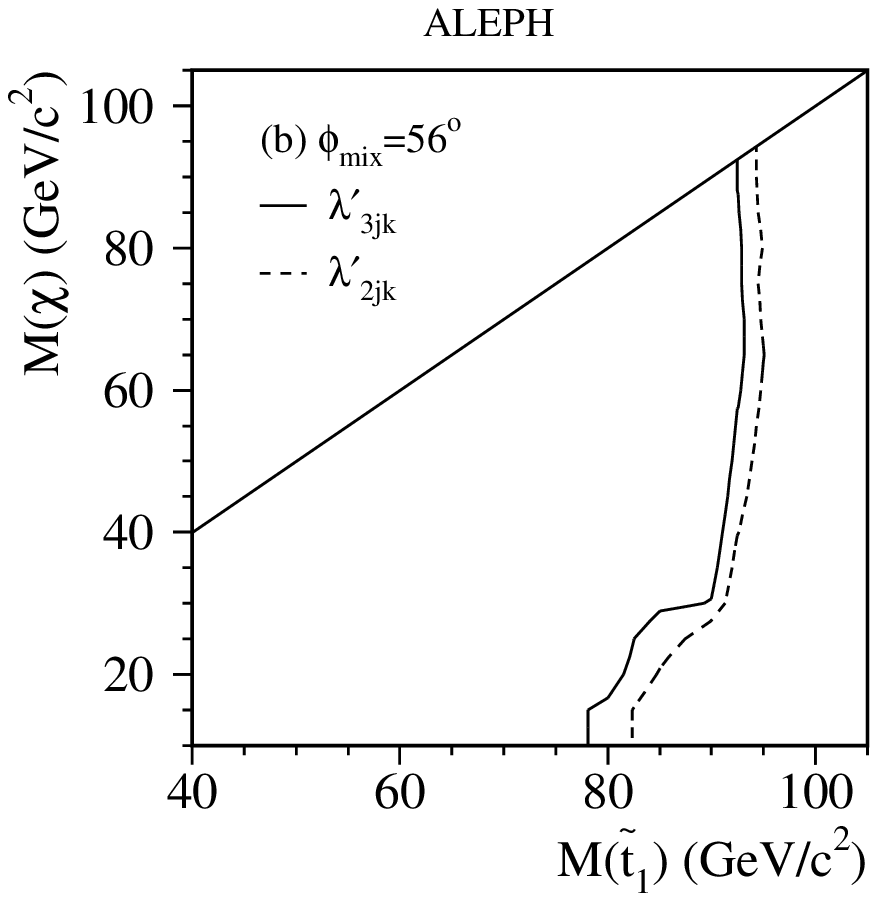}
}
\resizebox{\textwidth}{!}{
  \includegraphics{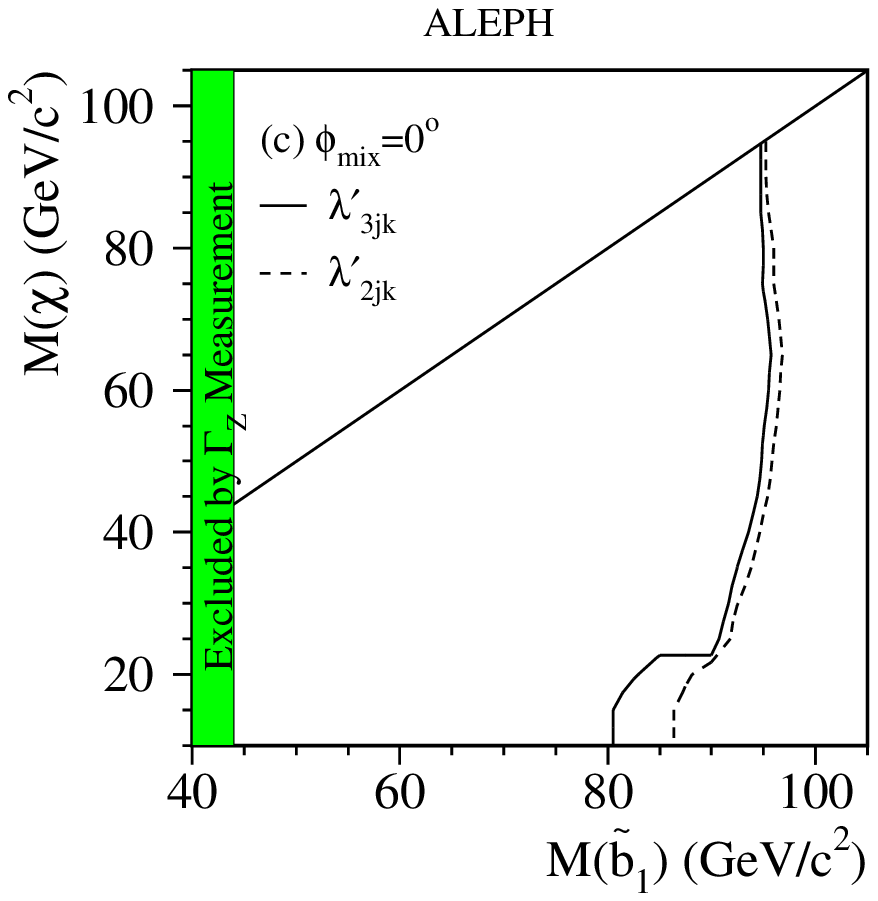}
  \includegraphics{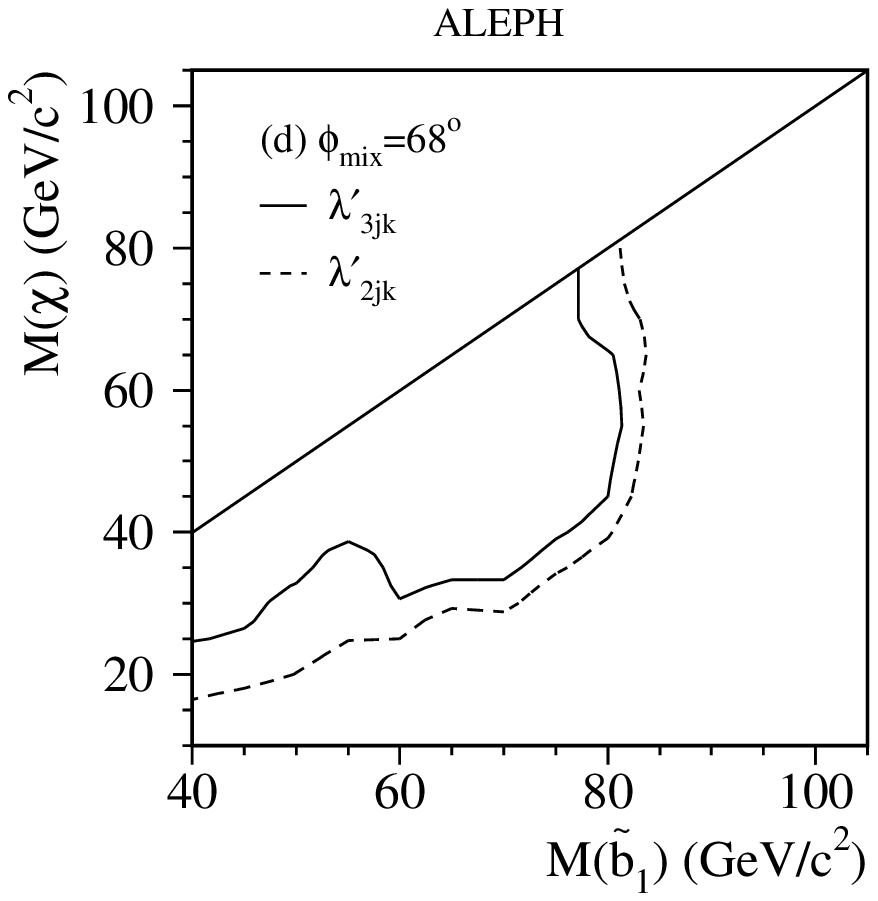}
}
\caption[.]{\label{fig:squark_lqd_ind.eps} 
{\small The $95\%$ C.L. limits in
(a),(b) the ($m_{\chi}$, $m_{\stone}$) plane and (c),(d) the ($m_{\chi}$, $m_{\sbone}$) 
plane for
indirect decays via a  
 $\lambda'_{211}$ or 
$\lambda'_{311}$ $\slqd$ coupling, 
for no mixing ($\phimix=0^\circ$) and for $\phimix=56^\circ$ and $68^\circ$
for stops and sbottoms, respectively.}}
\end{center}
\end{figure}

 Limits for  left-handed squarks are
shown in Fig.~\ref{fig:squark_lqd_ind.eps}. The following limits for
$\stL$ and $\sbL$ are derived: $m_{\stL}>85~\gevcc$ and
$m_{\sbL}>80~\gevcc$.

\subsection{Sleptons and Sneutrinos decaying via \boldmath{\slqd}}
\label{lqd.fourjets}
\begin{figure}
\begin{center}
\resizebox{\textwidth}{!}{
 \includegraphics{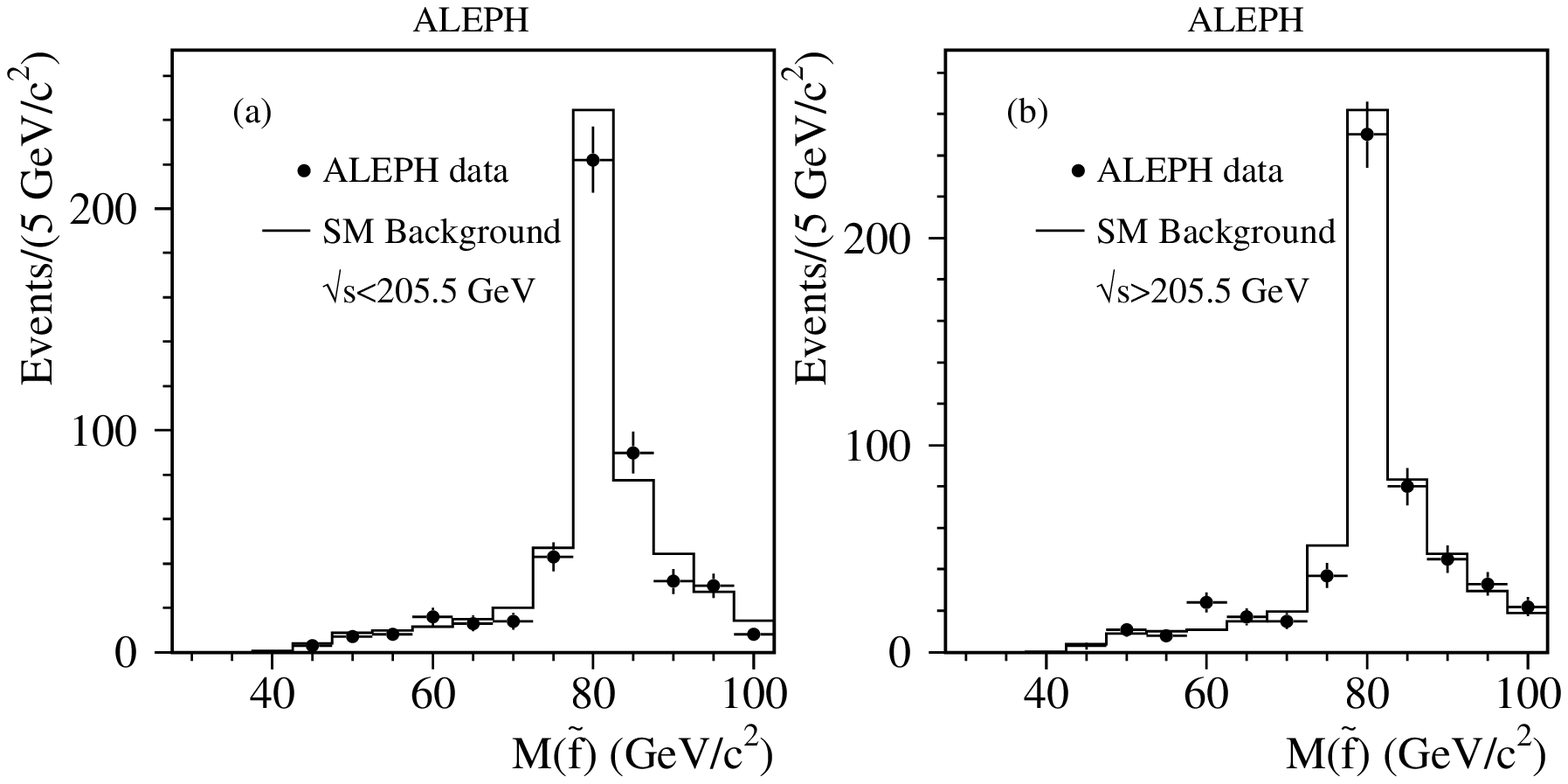}
}
\resizebox{0.5\textwidth}{!}{
 \includegraphics{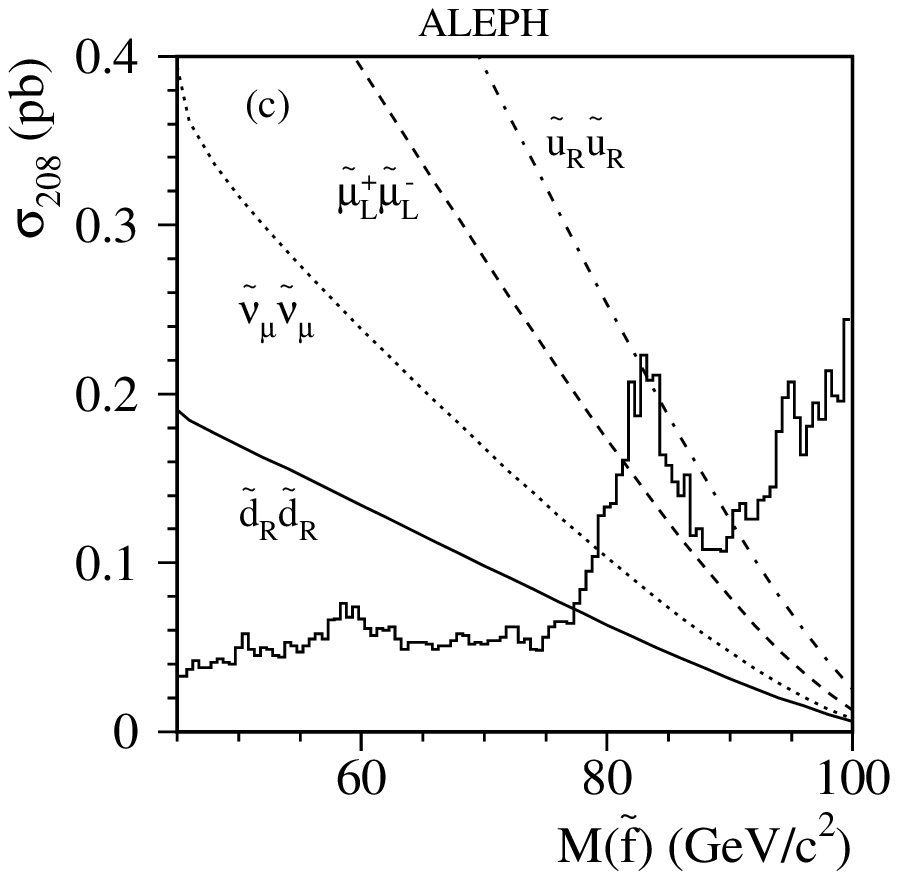}
}
\caption[.]{ \small \label{fig:fourjets}
The distributions of the reconstructed jet-pair invariant
masses after forcing each event into four jets. The
points are the data taken in year 2000,
for (a) the 205 \gev\ sample and (b)~the~207~\gev\ sample. 
The solid histogram is the predicted Standard Model
background.  In (c), the $95\%$ C.L. cross section upper limit
for sleptons (via $\slqd$), sneutrinos (via $\slqd$) and squarks (via
$\sudd$) decaying directly to four jets is shown. The MSSM cross sections for
pair production of muon sneutrinos, left-handed smuons and
right-handed squarks are superimposed.  }
\end{center}
\end{figure}

   Direct decays of sleptons and sneutrinos via the $\slqd$
operator lead to four-jet final states. The ``Four-Jets'' selection is
applied.  The distributions of the di-jet masses for data and Monte
Carlo are shown in Figs.~\ref{fig:fourjets}a and \ref{fig:fourjets}b. Limits are derived by
sliding a mass window of $5~\gevcc$ across the di-jet mass
distribution. The results are shown in Fig.~\ref{fig:fourjets}c and
imply $m_{\snumu}>79~\gevcc$ and $m_{\smuL}>81~\gevcc$.

\begin{figure}
\begin{center}
\resizebox{\textwidth}{!}{
  \includegraphics{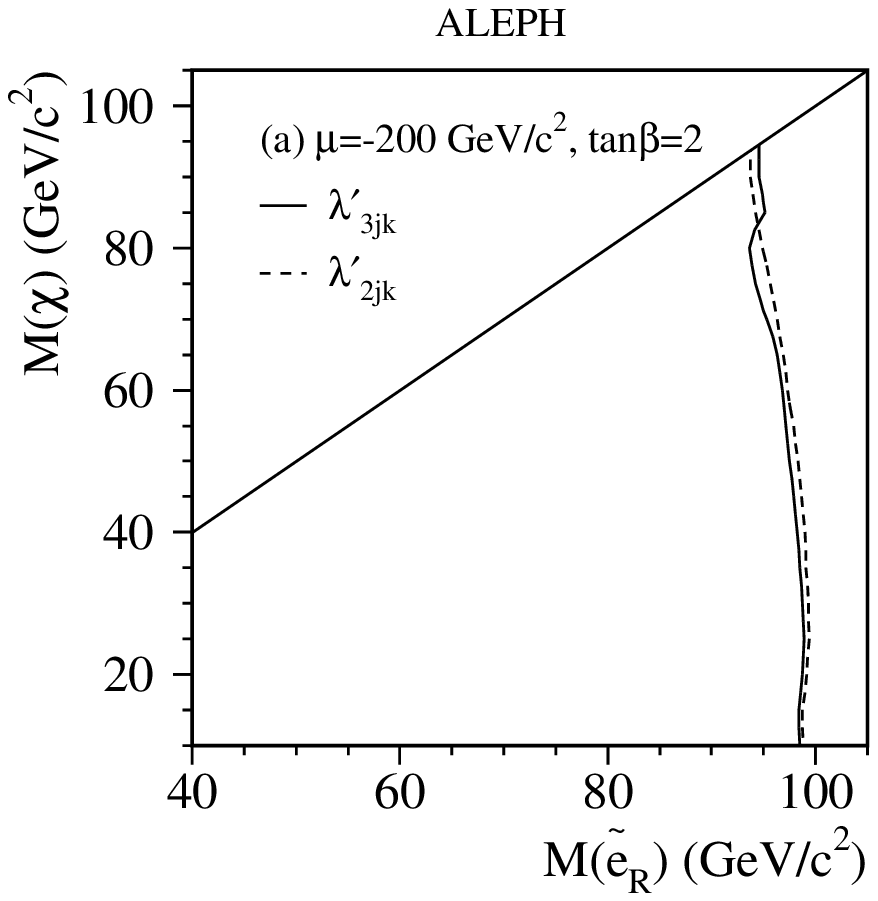}
  \includegraphics{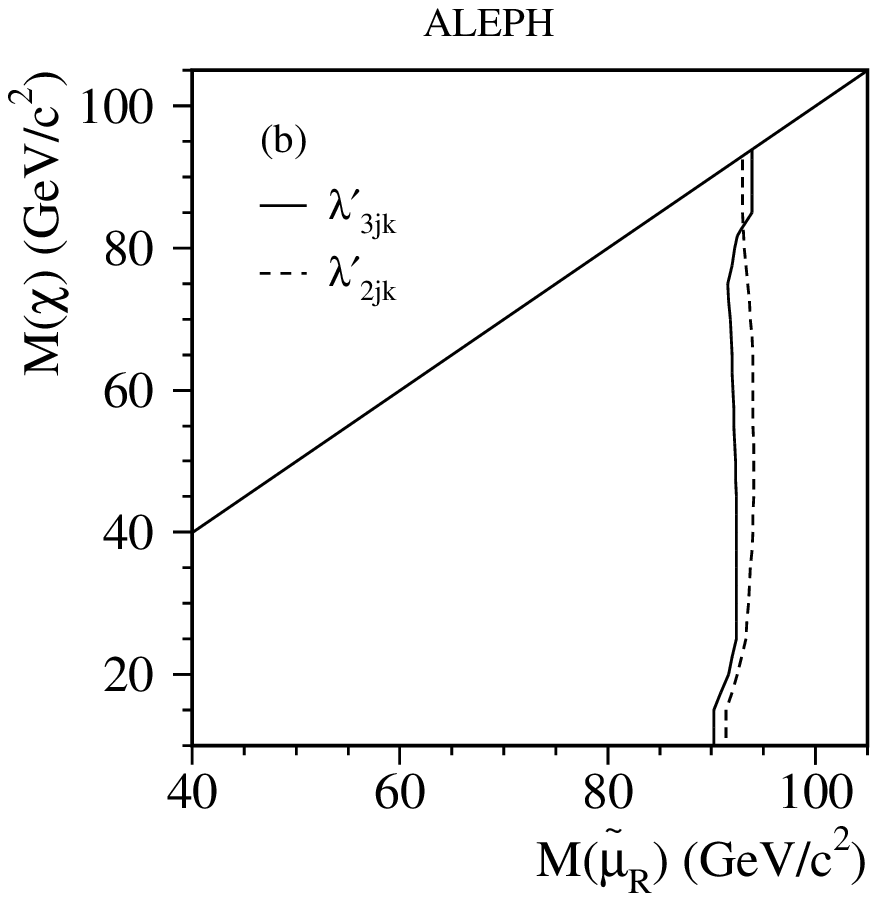}
}
\resizebox{0.5\textwidth}{!}{
  \includegraphics{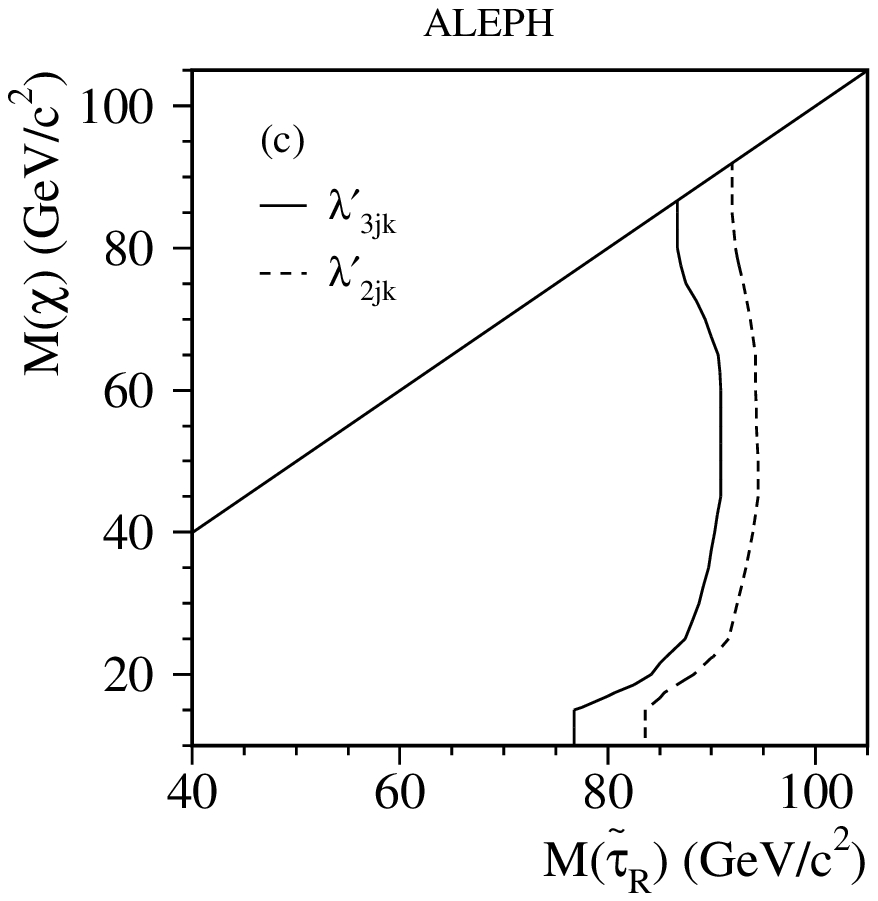}
}
\caption[.]{  \label{fig:slep_LQD_ind} \small
The $95\%$ C.L. limits in the ($m_{\chi}$, $m_{\slR}$)
plane for (a) selectrons, (b) smuons and (c) staus decaying indirectly via a
dominant $\slqd$ operator. The two choices of $\lambda'_{2jk}$ and
$\lambda'_{3jk}$ correspond to the most and least stringent exclusions,
respectively. The selectron cross section is evaluated 
at $\mu=-200~\gevcc$ and $\tanb=2$. The limit from the $\Gamma_\mathrm{Z}$
measurements excludes $m_{\slep}<38~\gevcc$. }
\end{center}
\end{figure}

\begin{figure}
\begin{center}
\resizebox{\textwidth}{!}{
  \includegraphics{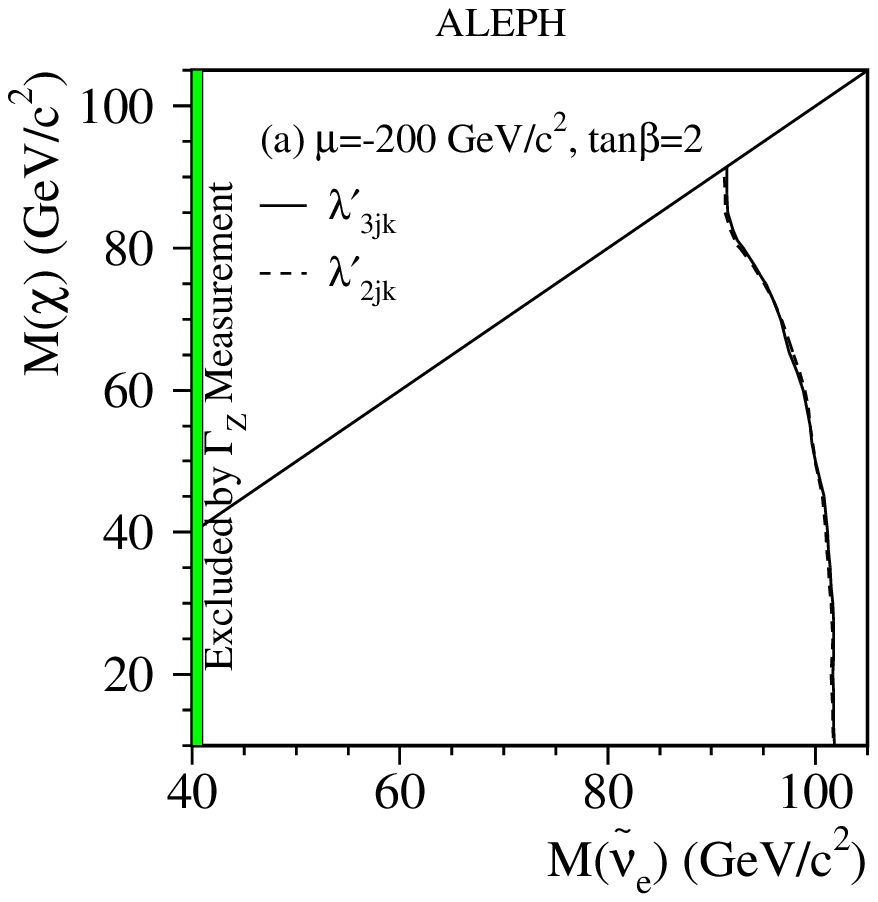}
  \includegraphics{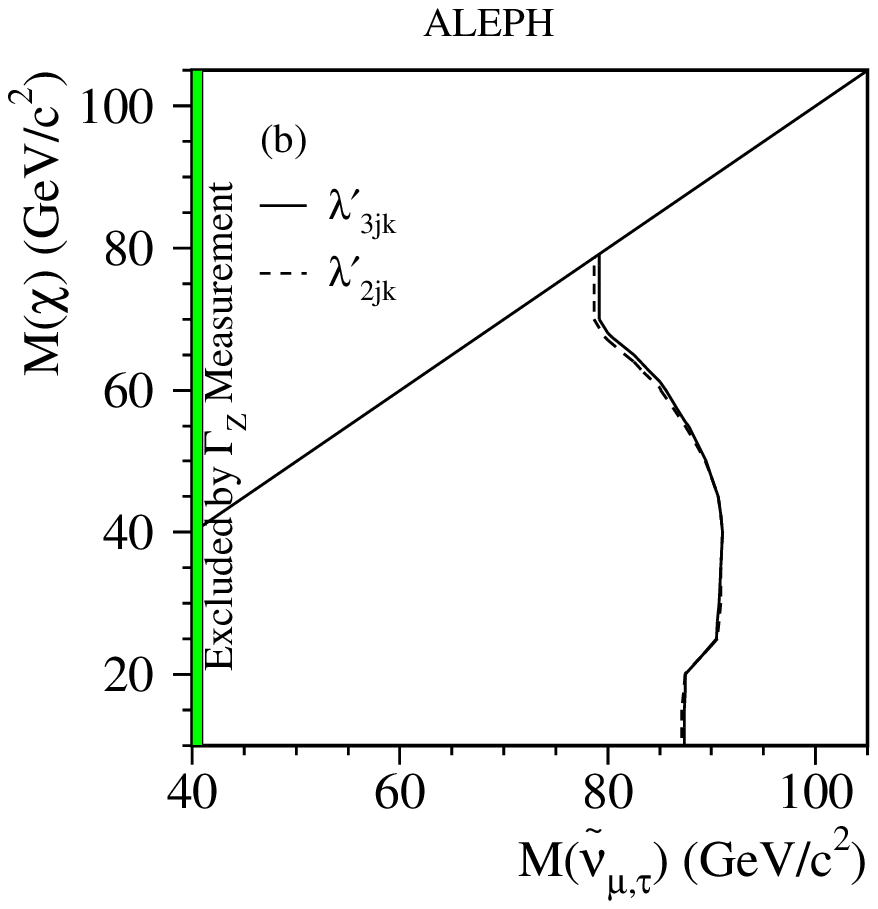}
}
\caption[.]{  \label{fig:snu_LQD_ind} \small
 The $95\%$ C.L. limits in the ($m_{\chi}$, $m_{\snu}$) plane for (a)
electron and (b) muon or tau sneutrinos decaying indirectly via a dominant
$\slqd$ operator. The two choices of $\lambda'_{2jk}$ and
$\lambda'_{3jk}$ correspond to the most and least stringent exclusions,
respectively. The electron sneutrino cross section is evaluated at
$\mu=-200~\gevcc$ and $\tanb=2$.}
\end{center}
\end{figure}

Indirect decays of the sleptons via the $\slqd$ operator yield
two, three or four leptons and four jets in the final state; two
leptons are of the same flavour as the initial sleptons. 
 The
indirect decays of sneutrinos produce a final state with four jets,
up to two leptons and missing energy. For selectrons and smuons the
``4 Jets + 2 Iso.~$\ell$'' selection is used except for the special case
of $\lambda'_{3jk}\neq0$ and $(m_{\slR}-m_\chi)<10~\gevcc$ where the
``4 Jets + 2$\tau$'' selection is used. Indirect stau decays are
selected with the ``5 Jets + 1 Iso.~$\ell$'' selection if
$m_\chi>20~\gevcc$ and either $\lambda'_{2jk}\neq0$ or
$\lambda'_{1jk}\neq0$. The combination of the ``5
Jets + 1 Iso.~$\ell$'' and the ``Leptons and Hadrons'' selections is
used otherwise. The sneutrinos are selected with the ``4JH'' selection for
$m_\chi>20~\gevcc$ and ``AJ-H'' (acoplanar jets) otherwise. Limits for
sleptons with indirect decays are shown in 
Figs.~\ref{fig:slep_LQD_ind}~and~\ref{fig:snu_LQD_ind} with
 the selectron
and electron sneutrino cross sections evaluated at
$\mu=-200~\gevcc$ and $\tanb=2$.
The limits are $m_{\seR}>93~\gevcc$,
$m_{\smuR}>90~\gevcc$, $m_{\stauR}>76~\gevcc$, $m_{\snue}>91~\gevcc$
and $m_{\snumu}>78~\gevcc$.

%\pagebreak

\subsection{Charginos and neutralinos decaying via \boldmath{\sudd}}
\label{udd.charginos}

The decay of the lightest neutralino leads to six hadronic jets in the final state.
The indirect decays of a chargino or the second lightest neutralino
give rise to a variety of final states which range from ten hadronic jets to six
jets associated with leptons and missing energy.  The ``Many Jets'',
``Four Jets'' and ``Many Jets + Lepton'' selections are used to cover
these topologies.

In the framework
of the MSSM, 
$95\%$ C.L. exclusion limits in the ($\mu$, $M_2$) plane are obtained as
shown in Fig.~\ref{mum2lims}c. 
 The
lower limit on the lightest chargino mass is  $103~\gevcc$.

The searches for the lightest and second lightest
neutralino
 do not extend the excluded region in the
($\mu$,$M_2$) plane beyond that achieved with the chargino 
search alone.

\subsection {Squarks decaying via \boldmath{\sudd}}

The direct decay of pair produced squarks leads to four-quark final
states. The ``Four Jet'' selection is therefore used to extract the
mass limits.  As shown in Fig.~\ref{fig:fourjets}c the mass limits
are $82.5~\gevcc$ for up-type squarks and $77~\gevcc$ for down-type
squarks.

\begin{figure}
\begin{center}
\resizebox{\textwidth}{!}{
 \includegraphics{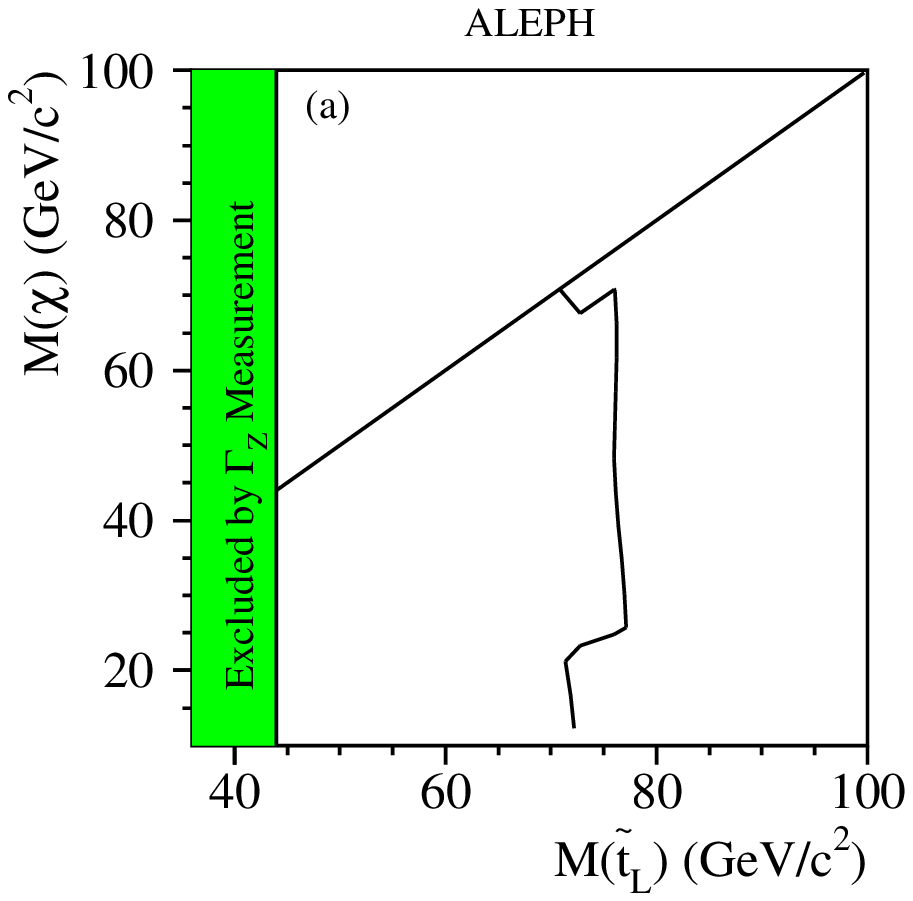}
 \includegraphics{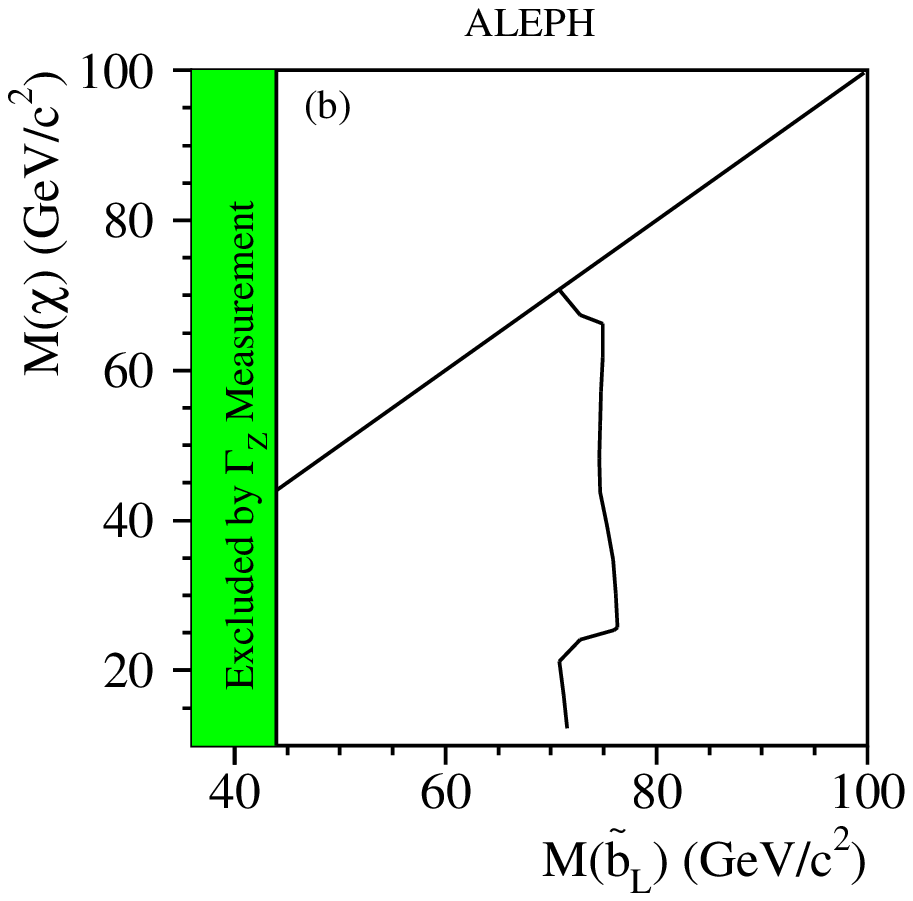}
}
\caption[.]{\label{udd_stop} 
{\small The $95\%$ C.L. limits in
(a) the ($m_{\chi}$, $m_{\stL}$) plane and (b) the ($m_{\chi}$, $m_{\sbL}$) 
plane for
indirect decays via the $\sudd$ couplings.}}
\end{center}
\end{figure}

For indirect squark decays, which lead to eight-jet
topologies, the ``Four Jets Broad'' and ``Four Jets'' selections are used. 
Figure~\ref{udd_stop} shows the $95\%$ C.L. exclusion limits 
in the ($m_{\chi}$, $m_{\sq}$) plane for
left-handed stop and sbottom.  The
corresponding mass limits are $m_{\stL}>71.5~\gevcc$ and
$m_{\sbL}>71.5~\gevcc$.

\subsection {Sleptons decaying via \boldmath{\sudd}}

No direct slepton decays are possible via the $\sudd$ coupling.  For
the indirect decays of pair produced selectrons and smuons, which lead to
six-jet plus two-lepton final states, the ``Four Jets + 2 Leptons''
selection is used for large mass differences between the slepton and
neutralino, and the ``Many Jets + 2 Leptons'' for the low mass
difference region. In addition, for the very low mass difference
region the leptons are very soft and the ``Four Jets'' selection is
used. For indirect stau decays, which lead to six-jet plus two tau final states,
 the ``Four Jets + Taus'' and ``Many Jets + Taus'' selections are used 
for large and low mass 
differences between the stau and the
neutralino, respectively.

\begin{figure}
\begin{center}
\resizebox{\textwidth}{!}{
  \includegraphics{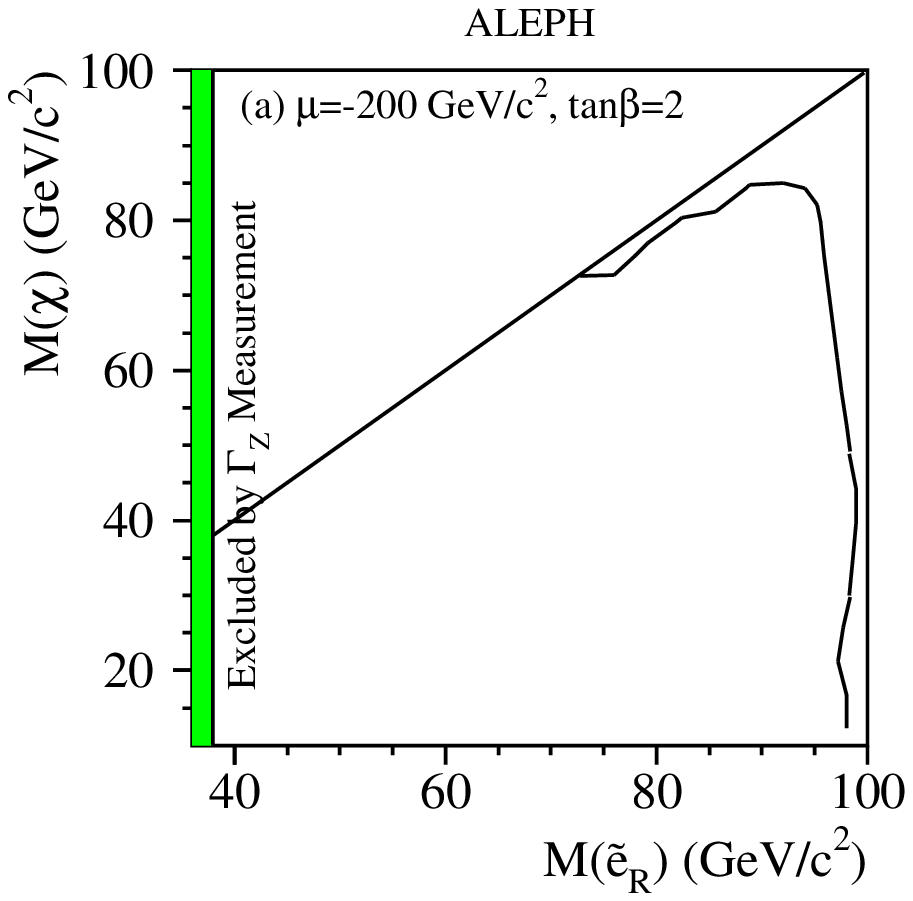}
  \includegraphics{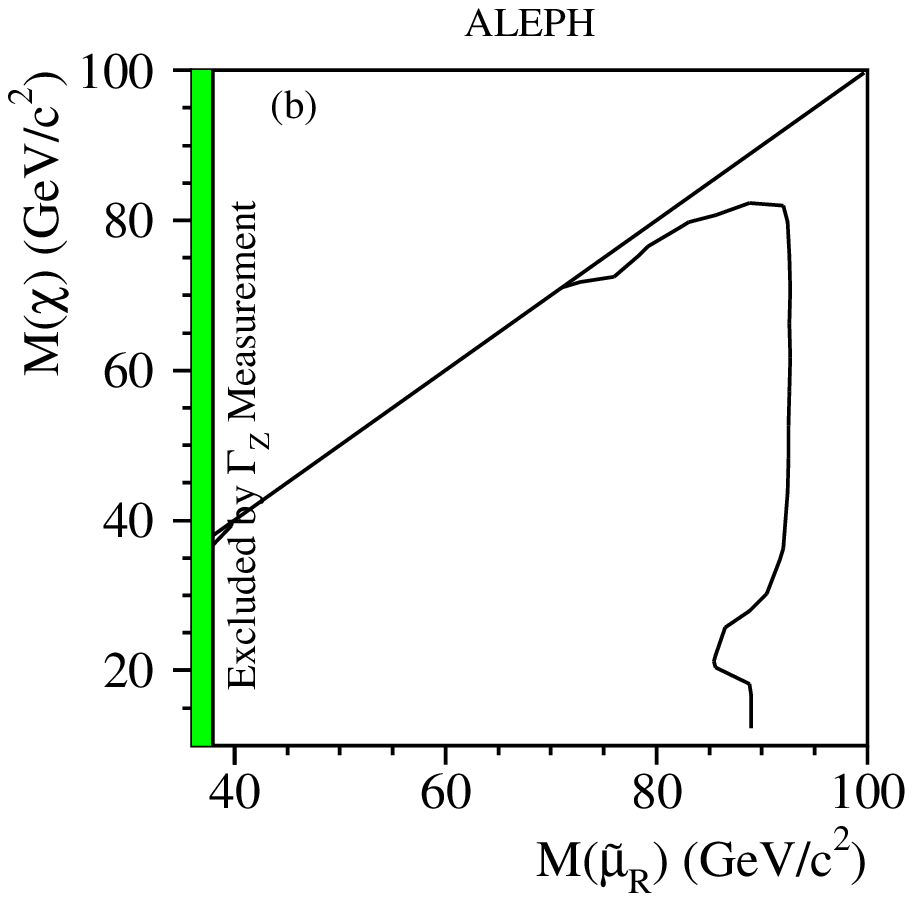}
}
\resizebox{0.5\textwidth}{!}{
 \includegraphics{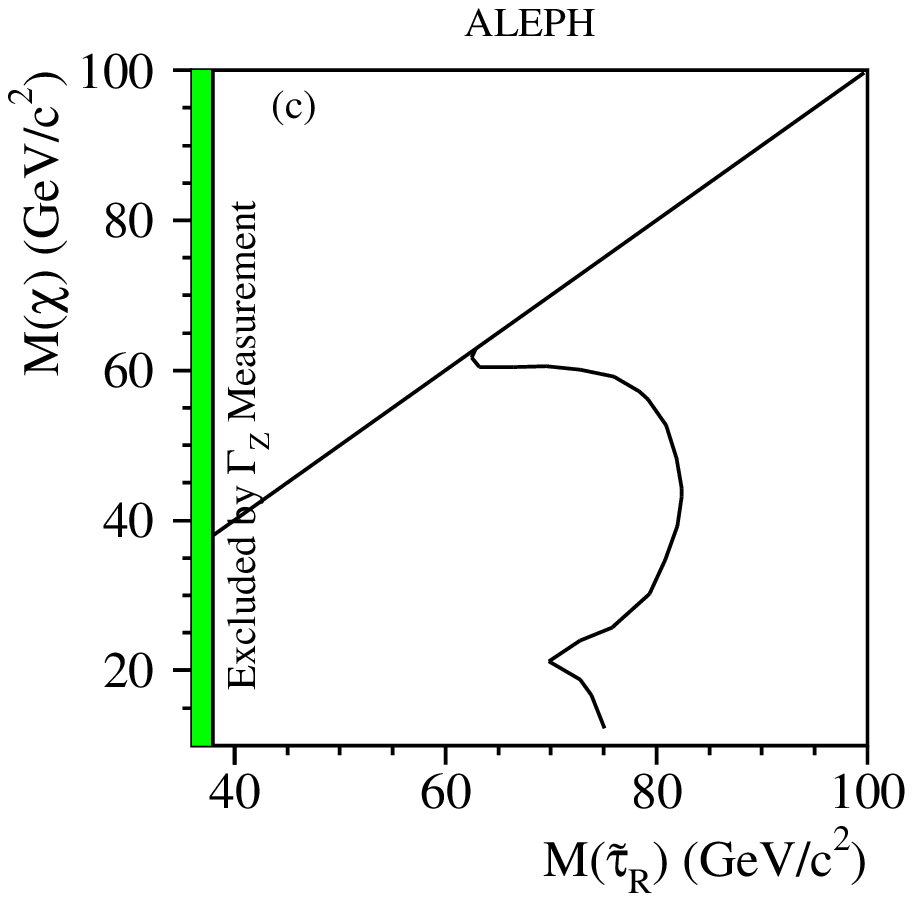}
}
\caption[.]{  \label{udd_slep}
{ \small
The $95\%$ C.L. limits in the ($m_{\chi}$, $m_{\slep}$)
plane for
(a) selectrons , (b) smuons and (c)~staus 
 decaying indirectly via a
dominant $\sudd$ operator.
  The selectron cross section is evaluated 
at $\mu=-200~\gevcc$ and $\tanb=2$.}}

\end{center}
\end{figure}

The resulting $95\%$ C.L. exclusion limits in the
($m_{\chi}$, $m_{\slep}$) plane are shown in 
 Fig.~\ref{udd_slep}.  The
selectron cross section is evaluated at $\mu=-200~\gevcc$ and
$\tanb=2$. For $m_{\slep}-m_{\chi}>10~\gevcc$ this yields
$m_{\seR}>94~\gevcc$, $m_{\smuR}>85~\gevcc$ and $m_{\stauR}>70~\gevcc$.

%\pagebreak
\subsection {Sneutrinos decaying via \boldmath{\sudd}  }
 \label{udd.sneutrinos}

No direct sneutrino decays are possible via the $\sudd$
coupling.  Sneutrinos decaying indirectly lead to six-jet final 
states plus two neutrinos. 
 For small mass differences between the sneutrino
and neutralino, the six jets are well separated and the 
``Many Jets + $\emiss$'' selection is used.
For large mass differences  the event is characterized by a significant missing
energy, and the ``Four Jets + $\emiss$'' selection is used. 

\begin{figure}
\begin{center}
\resizebox{\textwidth}{!}{
  \includegraphics{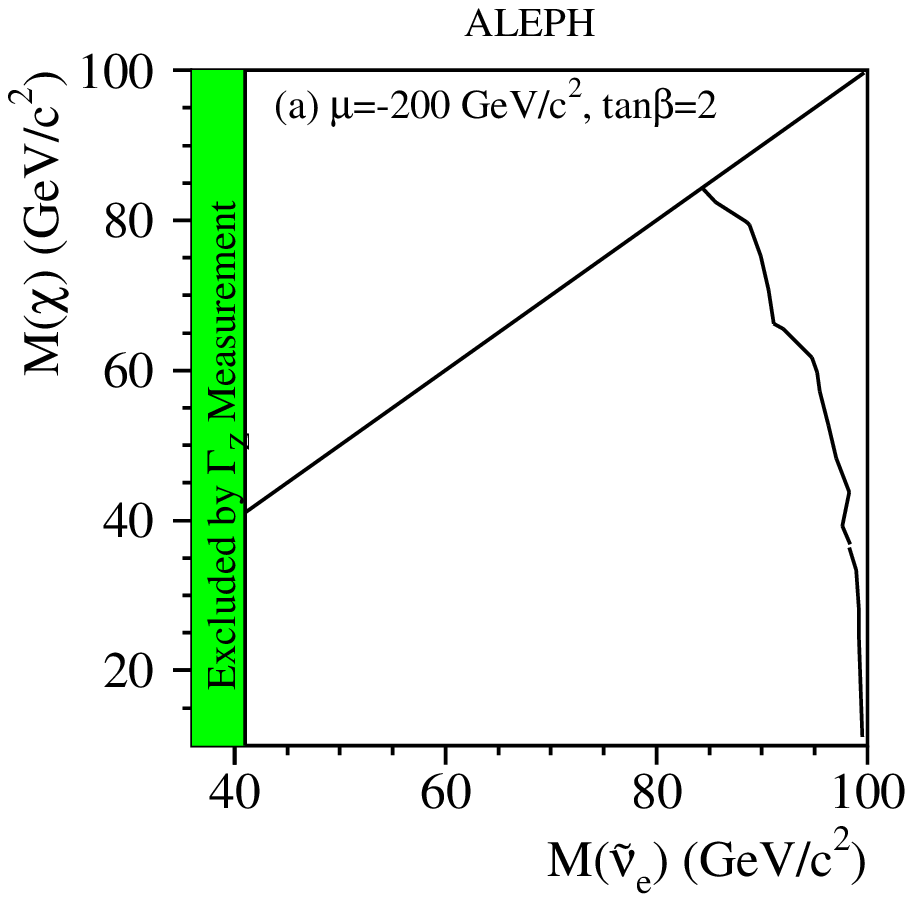}
  \includegraphics{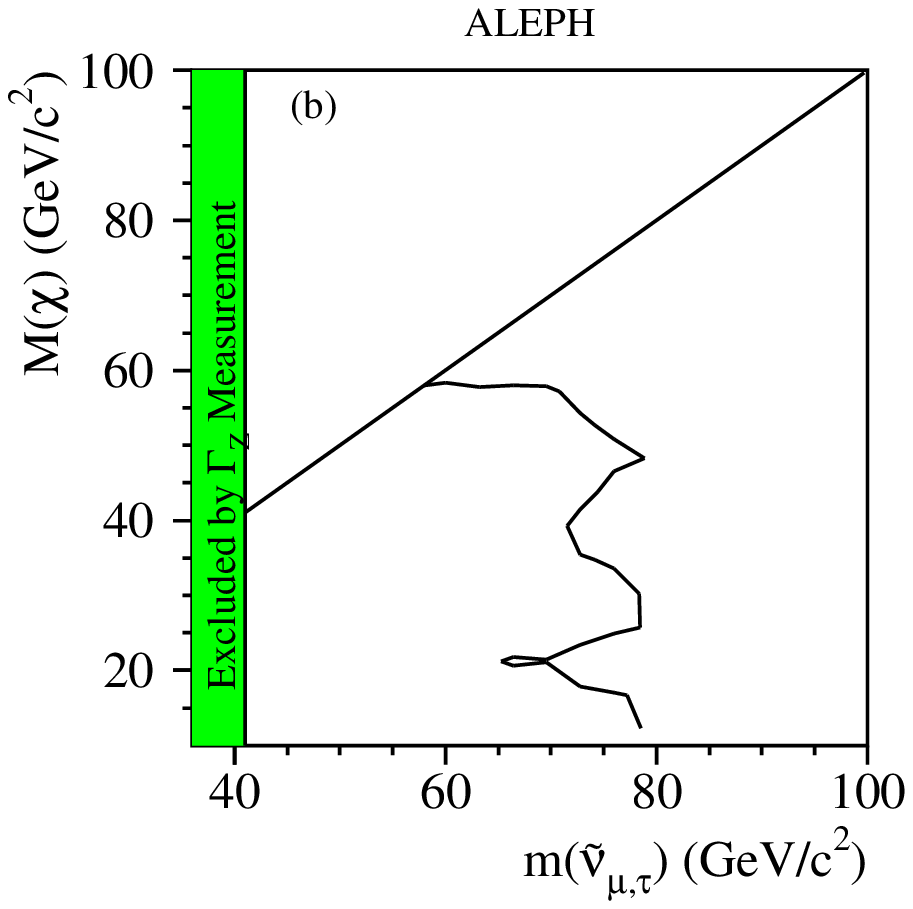}
}
\caption[.]{  \label{udd_sneu}{(a) \small The $95\%$ C.L. limits in the
($m_{\chi}$, $m_{\snu}$) plane for $\snu_\mathrm{e}$ decaying indirectly via a
dominant $\sudd$ operator.  The $\snu_\mathrm{e}$ cross section is evaluated at
$\mu=-200~\gevcc$ and $\tanb=2$. (b) The exclusion obtained in the
($m_{\chi}$, $m_{\snu_{\mu,\tau}}$) plane for $\snu_{\mu,\tau}$
decaying indirectly via a dominant $\sudd$ operator.}}
\end{center}
\end{figure}

The $95\%$ C.L. exclusions in the ($m_{\chi}$, $m_{\snu}$) plane are shown 
in Fig.~\ref{udd_sneu}a for the electron sneutrino and in 
Fig.~\ref{udd_sneu}b for the muon or tau sneutrino.
The electron sneutrino cross section is
evaluated at $\mu=-200~\gevcc$ and $\tanb=2$. The limits
$m_{\snue}>88~\gevcc$ and $m_{\snu_{\mu,\tau}}>65~\gevcc$ are
obtained for $m_{\snu}-m_{\chi}>10~\gevcc$ .

\section{Summary}\label{conclusions}

Pair production of supersymmetric particles, followed by
direct or indirect decays involving R-parity violating couplings,
has been searched for in the data collected with the ALEPH detector 
at LEP at centre-of-mass energies between 189 and 209 \gev.
It has been
assumed that the LSP has a negligible lifetime, and that only one
$\lambda_{ijk},\lambda'_{ijk}$ or $\lambda''_{ijk}$ coupling is
nonzero.
Several selections covering all the possible final states have been
applied.
No evidence for a signal has been found 
and various limits have been set within the framework of the MSSM with
R-parity violating couplings.
These results
improve on those previously
published by ALEPH~\cite{paper:202_rpv} and by the 
other LEP collaborations~\cite{OtherLEP}.

The limits obtained for direct decays of sfermions are
\begin{itemize}
\item for an $\slle$ coupling:
\begin{itemize}
\item[-] $m_{\seR}>96~\gevcc$ ($\mu=-200~\gevcc$, $\tanb=2$), 
\item[-] $m_{\snu_\e}>100~\gevcc$ ($\mu=-200~\gevcc$, $\tanb=2$),
\item[-] $m_{\smuR,\stauR}>87~\gevcc$,
\item[-] $m_{\snu_\mu}>90~\gevcc$. 
\end{itemize}

\item for an $\slqd$ coupling:  
\begin{itemize}
\item[-] $m_{\smuL}~>81~\gevcc$, 
\item[-] $m_{\snu_\mu}>79~\gevcc$,
\item[-] $m_{\stL}~>~97~\gevcc$ for Br$(\stL\ra q\tau)=1$.
\end{itemize}

\item for a $\sudd$ coupling:  
\begin{itemize}
\item[-] $m_{\suL}>82.5~\gevcc$,
\item[-] $m_{\sdL}>77.0~\gevcc$.
\end{itemize}

\end{itemize}

\begin{table}[hhh]
\caption[.]{ \small 
The $95\%$ confidence level lower mass limits for indirect sparticle decays
for each of the three R-parity violating couplings, assuming 
$m_{\slep,\snu}-m_{\chi}>10~\gevcc$ for $\sudd$ and $\mu=-200~\gevcc$ 
and $\tanb=2$ for $\se$ and $\snue$.
}
\label{table:indirect_limits}
\begin{center}
\renewcommand{\arraystretch}{1.1}
\begin{tabular}{|c|c|c|c|}
\hline
 & \multicolumn{3}{c|}{Lower mass limit ($\gevcc$)} \\
\cline{2-4} 
Sparticle &~~~$\slle$~~~ & ~~~$\slqd$~~~ & ~~~$\sudd$~~~ \\
\hline
$\stL$ & 91  & 85  & 71.5  \\
$\sbL$ & 90  & 80  & 71.5  \\
$\seR$   & 96  & 93  & 94  \\
$\smuR$  & 96  & 90  & 85  \\
$\stauR$ & 95  & 76  & 70 \\
$\snue$  & 98  & 91  & 88  \\
$\snu_{\mu,\tau}$ &  89  & 78  & 65 \\

\hline
\end{tabular}
\renewcommand{\arraystretch}{1.}
\end{center}
\end{table}

For the indirect decays of sfermions, mass limits are listed
in Table~\ref{table:indirect_limits}. 
For large sfermion masses, an absolute limit of 103~\gevcc\ has been set
on the chargino mass, irrespective of the R-parity violating
operator.

\section{Acknowledgements}
It is a pleasure to congratulate our colleagues from the accelerator divisions
for the successful operation of LEP at high energy. 
We would like to express our gratitude to the engineers and 
support people at our home institutes without whose dedicated help
this work would not have been possible. 
Those of us from non-member states wish to thank CERN for its hospitality
and support.

\clearpage

\end{document}

%% file: authb.tex
%------------------------------------------------------------------------
% authob12pt.tex
% authors' list for papers at LEP 1.5 and 2 energies
%-----------------------------------------------------------------------
\pagestyle{empty}
\newpage
\small
%
% remember the old settings
%
\newlength{\saveparskip}
\newlength{\savetextheight}
\newlength{\savetopmargin}
\newlength{\savetextwidth}
\newlength{\saveoddsidemargin}
\newlength{\savetopsep}
\setlength{\saveparskip}{\parskip}
\setlength{\savetextheight}{\textheight}
\setlength{\savetopmargin}{\topmargin}
\setlength{\savetextwidth}{\textwidth}
\setlength{\saveoddsidemargin}{\oddsidemargin}
\setlength{\savetopsep}{\topsep}
%
% text dimensions for the author list
%
\setlength{\parskip}{0.0cm}
\setlength{\textheight}{25.0cm}
\setlength{\topmargin}{-1.5cm}
\setlength{\textwidth}{16 cm}
\setlength{\oddsidemargin}{-0.0cm}
\setlength{\topsep}{1mm}
\pretolerance=10000
%%\begin{document}
%\centerline{EUROPEAN ORGANIZATION FOR NUCLEAR RESEARCH}
%\centerline{EUROPEAN LABORATORY FOR PARTICLE PHYSICS (CERN)}
%\vspace{1cm}
%\begin{flushright}CERN-EP-2000-
%\9 August 2002 - last update
%\end{flushright}
\centerline{\large\bf The ALEPH Collaboration}
\footnotesize
\vspace{0.5cm}
{\raggedbottom
\begin{sloppypar}
\samepage\noindent
A.~Heister,
S.~Schael
\nopagebreak
\begin{center}
\parbox{15.5cm}{\sl\samepage
Physikalisches Institut das RWTH-Aachen, D-52056 Aachen, Germany}
\end{center}\end{sloppypar}
\vspace{2mm}
\begin{sloppypar}
\noindent
R.~Barate,
R.~Bruneli\`ere,
I.~De~Bonis,
D.~Decamp,
C.~Goy,
S.~Jezequel,
J.-P.~Lees,
F.~Martin,
E.~Merle,
\mbox{M.-N.~Minard},
B.~Pietrzyk,
B.~Trocm\'e
\nopagebreak
\begin{center}
\parbox{15.5cm}{\sl\samepage
Laboratoire de Physique des Particules (LAPP), IN$^{2}$P$^{3}$-CNRS,
F-74019 Annecy-le-Vieux Cedex, France}
\end{center}\end{sloppypar}
\vspace{2mm}
\begin{sloppypar}
\noindent
S.~Bravo,
M.P.~Casado,
M.~Chmeissani,
J.M.~Crespo,
E.~Fernandez,
M.~Fernandez-Bosman,
Ll.~Garrido,$^{15}$
M.~Martinez,
A.~Pacheco,
H.~Ruiz
\nopagebreak
\begin{center}
\parbox{15.5cm}{\sl\samepage
Institut de F\'{i}sica d'Altes Energies, Universitat Aut\`{o}noma
de Barcelona, E-08193 Bellaterra (Barcelona), Spain$^{7}$}
\end{center}\end{sloppypar}
\vspace{2mm}
\begin{sloppypar}
\noindent
A.~Colaleo,
D.~Creanza,
N.~De~Filippis,
M.~de~Palma,
G.~Iaselli,
G.~Maggi,
M.~Maggi,
S.~Nuzzo,
A.~Ranieri,
G.~Raso,$^{24}$
F.~Ruggieri,
G.~Selvaggi,
L.~Silvestris,
P.~Tempesta,
A.~Tricomi,$^{3}$
G.~Zito
\nopagebreak
\begin{center}
\parbox{15.5cm}{\sl\samepage
Dipartimento di Fisica, INFN Sezione di Bari, I-70126 Bari, Italy}
\end{center}\end{sloppypar}
\vspace{2mm}
\begin{sloppypar}
\noindent
X.~Huang,
J.~Lin,
Q. Ouyang,
T.~Wang,
Y.~Xie,
R.~Xu,
S.~Xue,
J.~Zhang,
L.~Zhang,
W.~Zhao
\nopagebreak
\begin{center}
\parbox{15.5cm}{\sl\samepage
Institute of High Energy Physics, Academia Sinica, Beijing, The People's
Republic of China$^{8}$}
\end{center}\end{sloppypar}
\vspace{2mm}
\begin{sloppypar}
\noindent
D.~Abbaneo,
P.~Azzurri,
T.~Barklow,$^{26}$
O.~Buchm\"uller,$^{26}$
M.~Cattaneo,
F.~Cerutti,
B.~Clerbaux,$^{23}$
H.~Drevermann,
R.W.~Forty,
M.~Frank,
F.~Gianotti,
J.B.~Hansen,
J.~Harvey,
D.E.~Hutchcroft,
P.~Janot,
B.~Jost,
M.~Kado,$^{2}$
P.~Mato,
A.~Moutoussi,
F.~Ranjard,
L.~Rolandi,
D.~Schlatter,
G.~Sguazzoni,
W.~Tejessy,
F.~Teubert,
A.~Valassi,
I.~Videau,
J.J.~Ward
\nopagebreak
\begin{center}
\parbox{15.5cm}{\sl\samepage
European Laboratory for Particle Physics (CERN), CH-1211 Geneva 23,
Switzerland}
\end{center}\end{sloppypar}
\vspace{2mm}
\begin{sloppypar}
\noindent
F.~Badaud,
S.~Dessagne,
A.~Falvard,$^{20}$
D.~Fayolle,
P.~Gay,
J.~Jousset,
B.~Michel,
S.~Monteil,
D.~Pallin,
J.M.~Pascolo,
P.~Perret
\nopagebreak
\begin{center}
\parbox{15.5cm}{\sl\samepage
Laboratoire de Physique Corpusculaire, Universit\'e Blaise Pascal,
IN$^{2}$P$^{3}$-CNRS, Clermont-Ferrand, F-63177 Aubi\`{e}re, France}
\end{center}\end{sloppypar}
\vspace{2mm}
\begin{sloppypar}
\noindent
J.D.~Hansen,
J.R.~Hansen,
P.H.~Hansen,
B.S.~Nilsson
\nopagebreak
\begin{center}
\parbox{15.5cm}{\sl\samepage
Niels Bohr Institute, 2100 Copenhagen, DK-Denmark$^{9}$}
\end{center}\end{sloppypar}
\vspace{2mm}
\begin{sloppypar}
\noindent
A.~Kyriakis,
C.~Markou,
E.~Simopoulou,
A.~Vayaki,
K.~Zachariadou
\nopagebreak
\begin{center}
\parbox{15.5cm}{\sl\samepage
Nuclear Research Center Demokritos (NRCD), GR-15310 Attiki, Greece}
\end{center}\end{sloppypar}
\vspace{2mm}
\begin{sloppypar}
\noindent
A.~Blondel,$^{12}$
\mbox{J.-C.~Brient},
F.~Machefert,
A.~Roug\'{e},
M.~Swynghedauw,
R.~Tanaka
\linebreak
H.~Videau
\nopagebreak
\begin{center}
\parbox{15.5cm}{\sl\samepage
Laoratoire Leprince-Ringuet, Ecole
Polytechnique, IN$^{2}$P$^{3}$-CNRS, \mbox{F-91128} Palaiseau Cedex, France}
\end{center}\end{sloppypar}
\vspace{2mm}
\begin{sloppypar}
\noindent
V.~Ciulli,
E.~Focardi,
G.~Parrini
\nopagebreak
\begin{center}
\parbox{15.5cm}{\sl\samepage
Dipartimento di Fisica, Universit\`a di Firenze, INFN Sezione di Firenze,
I-50125 Firenze, Italy}
\end{center}\end{sloppypar}
\vspace{2mm}
\begin{sloppypar}
\noindent
A.~Antonelli,
M.~Antonelli,
G.~Bencivenni,
F.~Bossi,
G.~Capon,
V.~Chiarella,
P.~Laurelli,
G.~Mannocchi,$^{5}$
G.P.~Murtas,
L.~Passalacqua
\nopagebreak
\begin{center}
\parbox{15.5cm}{\sl\samepage
Laboratori Nazionali dell'INFN (LNF-INFN), I-00044 Frascati, Italy}
\end{center}\end{sloppypar}
\vspace{2mm}
%\pagebreak
\begin{sloppypar}
\noindent
J.~Kennedy,
J.G.~Lynch,
P.~Negus,
V.~O'Shea,
A.S.~Thompson
\nopagebreak
\begin{center}
\parbox{15.5cm}{\sl\samepage
Department of Physics and Astronomy, University of Glasgow, Glasgow G12
8QQ,United Kingdom$^{10}$}
\end{center}\end{sloppypar}
\vspace{2mm}
%\pagebreak
\begin{sloppypar}
\noindent
S.~Wasserbaech
\nopagebreak
\begin{center}
\parbox{15.5cm}{\sl\samepage
Department of Physics, Haverford College, Haverford, PA 19041-1392, U.S.A.}
\end{center}\end{sloppypar}
\vspace{2mm}
%\pagebreak
\begin{sloppypar}
\noindent
R.~Cavanaugh,$^{4}$
S.~Dhamotharan,$^{21}$
C.~Geweniger,
P.~Hanke,
V.~Hepp,
E.E.~Kluge,
G.~Leibenguth,
A.~Putzer,
H.~Stenzel,
K.~Tittel,
M.~Wunsch$^{19}$
\nopagebreak
\begin{center}
\parbox{15.5cm}{\sl\samepage
Kirchhoff-Institut f\"ur Physik, Universit\"at Heidelberg, D-69120
Heidelberg, Germany$^{16}$}
\end{center}\end{sloppypar}
\vspace{2mm}
\begin{sloppypar}
\noindent
R.~Beuselinck,
W.~Cameron,
G.~Davies,
P.J.~Dornan,
M.~Girone,$^{1}$
R.D.~Hill,
N.~Marinelli,
J.~Nowell,
S.A.~Rutherford,
J.K.~Sedgbeer,
J.C.~Thompson,$^{14}$
R.~White
\nopagebreak
\begin{center}
\parbox{15.5cm}{\sl\samepage
Department of Physics, Imperial College, London SW7 2BZ,
United Kingdom$^{10}$}
\end{center}\end{sloppypar}
\vspace{2mm}
\begin{sloppypar}
\noindent
V.M.~Ghete,
P.~Girtler,
E.~Kneringer,
D.~Kuhn,
G.~Rudolph
\nopagebreak
\begin{center}
\parbox{15.5cm}{\sl\samepage
Institut f\"ur Experimentalphysik, Universit\"at Innsbruck, A-6020
Innsbruck, Austria$^{18}$}
\end{center}\end{sloppypar}
\vspace{2mm}
\begin{sloppypar}
\noindent
E.~Bouhova-Thacker,
C.K.~Bowdery,
D.P.~Clarke,
G.~Ellis,
A.J.~Finch,
F.~Foster,
G.~Hughes,
R.W.L.~Jones,
M.R.~Pearson,
N.A.~Robertson,
M.~Smizanska
\nopagebreak
\begin{center}
\parbox{15.5cm}{\sl\samepage
Department of Physics, University of Lancaster, Lancaster LA1 4YB,
United Kingdom$^{10}$}
\end{center}\end{sloppypar}
\vspace{2mm}
\begin{sloppypar}
\noindent
O.~van~der~Aa,
C.~Delaere,
V.~Lemaitre
\nopagebreak
\begin{center}
\parbox{15.5cm}{\sl\samepage
Institut de Physique Nucl\'eaire, D\'epartement de Physique, Universit\'e Catholique de Louvain, 1348 Louvain-la-Neuve, Belgium}
\end{center}\end{sloppypar}
\vspace{2mm}
\begin{sloppypar}
\noindent
U.~Blumenschein,
F.~H\"olldorfer,
K.~Jakobs,
F.~Kayser,
K.~Kleinknecht,
A.-S.~M\"uller,
B.~Renk,
H.-G.~Sander,
S.~Schmeling,
H.~Wachsmuth,
C.~Zeitnitz,
T.~Ziegler
\nopagebreak
\begin{center}
\parbox{15.5cm}{\sl\samepage
Institut f\"ur Physik, Universit\"at Mainz, D-55099 Mainz, Germany$^{16}$}
\end{center}\end{sloppypar}
\vspace{2mm}
\begin{sloppypar}
\noindent
A.~Bonissent,
P.~Coyle,
C.~Curtil,
A.~Ealet,
D.~Fouchez,
P.~Payre,
A.~Tilquin
\nopagebreak
\begin{center}
\parbox{15.5cm}{\sl\samepage
Centre de Physique des Particules de Marseille, Univ M\'editerran\'ee,
IN$^{2}$P$^{3}$-CNRS, F-13288 Marseille, France}
\end{center}\end{sloppypar}
\vspace{2mm}
\begin{sloppypar}
\noindent
F.~Ragusa
\nopagebreak
\begin{center}
\parbox{15.5cm}{\sl\samepage
Dipartimento di Fisica, Universit\`a di Milano e INFN Sezione di
Milano, I-20133 Milano, Italy.}
\end{center}\end{sloppypar}
\vspace{2mm}
\begin{sloppypar}
\noindent
A.~David,
H.~Dietl,
G.~Ganis,$^{27}$
K.~H\"uttmann,
G.~L\"utjens,
W.~M\"anner,
\mbox{H.-G.~Moser},
R.~Settles,
G.~Wolf
\nopagebreak
\begin{center}
\parbox{15.5cm}{\sl\samepage
Max-Planck-Institut f\"ur Physik, Werner-Heisenberg-Institut,
D-80805 M\"unchen, Germany\footnotemark[16]}
\end{center}\end{sloppypar}
\vspace{2mm}
\begin{sloppypar}
\noindent
J.~Boucrot,
O.~Callot,
M.~Davier,
L.~Duflot,
\mbox{J.-F.~Grivaz},
Ph.~Heusse,
A.~Jacholkowska,$^{6}$
L.~Serin,
\mbox{J.-J.~Veillet},
C.~Yuan
\nopagebreak
\begin{center}
\parbox{15.5cm}{\sl\samepage
Laboratoire de l'Acc\'el\'erateur Lin\'eaire, Universit\'e de Paris-Sud,
IN$^{2}$P$^{3}$-CNRS, F-91898 Orsay Cedex, France}
\end{center}\end{sloppypar}
\vspace{2mm}
\begin{sloppypar}
\noindent
%\samepage
G.~Bagliesi,
T.~Boccali,
L.~Fo\`a,
A.~Giammanco,
A.~Giassi,
F.~Ligabue,
A.~Messineo,
F.~Palla,
G.~Sanguinetti,
A.~Sciab\`a,
R.~Tenchini,$^{1}$
A.~Venturi,$^{1}$
P.G.~Verdini
\samepage
\begin{center}
\parbox{15.5cm}{\sl\samepage
Dipartimento di Fisica dell'Universit\`a, INFN Sezione di Pisa,
e Scuola Normale Superiore, I-56010 Pisa, Italy}
\end{center}\end{sloppypar}
\vspace{2mm}
\begin{sloppypar}
\noindent
O.~Awunor,
G.A.~Blair,
G.~Cowan,
A.~Garcia-Bellido,
M.G.~Green,
L.T.~Jones,
T.~Medcalf,
A.~Misiejuk,
J.A.~Strong,
P.~Teixeira-Dias
\nopagebreak
\begin{center}
\parbox{15.5cm}{\sl\samepage
Department of Physics, Royal Holloway \& Bedford New College,
University of London, Egham, Surrey TW20 OEX, United Kingdom$^{10}$}
\end{center}\end{sloppypar}
\vspace{2mm}
\begin{sloppypar}
\noindent
R.W.~Clifft,
T.R.~Edgecock,
P.R.~Norton,
I.R.~Tomalin
\nopagebreak
\begin{center}
\parbox{15.5cm}{\sl\samepage
Particle Physics Dept., Rutherford Appleton Laboratory,
Chilton, Didcot, Oxon OX11 OQX, United Kingdom$^{10}$}
\end{center}\end{sloppypar}
\vspace{2mm}
%\pagebreak
\begin{sloppypar}
\noindent
\mbox{B.~Bloch-Devaux},
D.~Boumediene,
P.~Colas,
B.~Fabbro,
E.~Lan\c{c}on,
\mbox{M.-C.~Lemaire},
E.~Locci,
P.~Perez,
J.~Rander,
B.~Tuchming,
B.~Vallage
\nopagebreak
\begin{center}
\parbox{15.5cm}{\sl\samepage
CEA, DAPNIA/Service de Physique des Particules,
CE-Saclay, F-91191 Gif-sur-Yvette Cedex, France$^{17}$}
\end{center}\end{sloppypar}
%\nopagebreak
\vspace{2mm}
\begin{sloppypar}
\noindent
N.~Konstantinidis,
A.M.~Litke,
G.~Taylor
\nopagebreak
\begin{center}
\parbox{15.5cm}{\sl\samepage
Institute for Particle Physics, University of California at
Santa Cruz, Santa Cruz, CA 95064, USA$^{22}$}
\end{center}\end{sloppypar}
%\pagebreak
\vspace{2mm}
\begin{sloppypar}
\noindent
C.N.~Booth,
S.~Cartwright,
F.~Combley,$^{25}$
P.N.~Hodgson,
M.~Lehto,
L.F.~Thompson
\nopagebreak
\begin{center}
\parbox{15.5cm}{\sl\samepage
Department of Physics, University of Sheffield, Sheffield S3 7RH,
United Kingdom$^{10}$}
\end{center}\end{sloppypar}
\vspace{2mm}
\begin{sloppypar}
\noindent
A.~B\"ohrer,
S.~Brandt,
C.~Grupen,
J.~Hess,
A.~Ngac,
G.~Prange
\nopagebreak
\begin{center}
\parbox{15.5cm}{\sl\samepage
Fachbereich Physik, Universit\"at Siegen, D-57068 Siegen, Germany$^{16}$}
\end{center}\end{sloppypar}
\vspace{2mm}
\begin{sloppypar}
\noindent
C.~Borean,
G.~Giannini
\nopagebreak
\begin{center}
\parbox{15.5cm}{\sl\samepage
Dipartimento di Fisica, Universit\`a di Trieste e INFN Sezione di Trieste,
I-34127 Trieste, Italy}
\end{center}\end{sloppypar}
\vspace{2mm}
\begin{sloppypar}
\noindent
H.~He,
J.~Putz,
J.~Rothberg
\nopagebreak
\begin{center}
\parbox{15.5cm}{\sl\samepage
Experimental Elementary Particle Physics, University of Washington, Seattle,
WA 98195 U.S.A.}
\end{center}\end{sloppypar}
\vspace{2mm}
\begin{sloppypar}
\noindent
S.R.~Armstrong,
K.~Berkelman,
K.~Cranmer,
D.P.S.~Ferguson,
Y.~Gao,$^{13}$
S.~Gonz\'{a}lez,
O.J.~Hayes,
H.~Hu,
S.~Jin,
J.~Kile,
P.A.~McNamara III,
J.~Nielsen,
Y.B.~Pan,
\mbox{J.H.~von~Wimmersperg-Toeller}, 
W.~Wiedenmann,
J.~Wu,
Sau~Lan~Wu,
X.~Wu,
G.~Zobernig
\nopagebreak
\begin{center}
\parbox{15.5cm}{\sl\samepage
Department of Physics, University of Wisconsin, Madison, WI 53706,
USA$^{11}$}
\end{center}\end{sloppypar}
\vspace{2mm}
\begin{sloppypar}
\noindent
G.~Dissertori
\nopagebreak
\begin{center}
\parbox{15.5cm}{\sl\samepage
Institute for Particle Physics, ETH H\"onggerberg, 8093 Z\"urich,
Switzerland.}
\end{center}\end{sloppypar}
}
\footnotetext[1]{Also at CERN, 1211 Geneva 23, Switzerland.}
\footnotetext[2]{Now at Fermilab, PO Box 500, MS 352, Batavia, IL 60510, USA}
\footnotetext[3]{Also at Dipartimento di Fisica di Catania and INFN Sezione di
 Catania, 95129 Catania, Italy.}
\footnotetext[4]{Now at University of Florida, Department of Physics, Gainesville, Florida 32611-8440, USA}
\footnotetext[5]{Also Istituto di Cosmo-Geofisica del C.N.R., Torino,
Italy.}
\footnotetext[6]{Also at Groupe d'Astroparticules de Montpellier, Universit\'{e} de Montpellier II, 34095, Montpellier, France.}
\footnotetext[7]{Supported by CICYT, Spain.}
\footnotetext[8]{Supported by the National Science Foundation of China.}
\footnotetext[9]{Supported by the Danish Natural Science Research Council.}
\footnotetext[10]{Supported by the UK Particle Physics and Astronomy Research
Council.}
\footnotetext[11]{Supported by the US Department of Energy, grant
DE-FG0295-ER40896.}
\footnotetext[12]{Now at Departement de Physique Corpusculaire, Universit\'e de
Gen\`eve, 1211 Gen\`eve 4, Switzerland.}
\footnotetext[13]{Also at Department of Physics, Tsinghua University, Beijing, The People's Republic of China.}
\footnotetext[14]{Supported by the Leverhulme Trust.}
\footnotetext[15]{Permanent address: Universitat de Barcelona, 08208 Barcelona,
Spain.}
\footnotetext[16]{Supported by Bundesministerium f\"ur Bildung
und Forschung, Germany.}
\footnotetext[17]{Supported by the Direction des Sciences de la
Mati\`ere, C.E.A.}
\footnotetext[18]{Supported by the Austrian Ministry for Science and Transport.}
\footnotetext[19]{Now at SAP AG, 69185 Walldorf, Germany}
\footnotetext[20]{Now at Groupe d' Astroparticules de Montpellier, Universit\'e de Montpellier II, 34095 Montpellier, France.}
\footnotetext[21]{Now at BNP Paribas, 60325 Frankfurt am Mainz, Germany}
\footnotetext[22]{Supported by the US Department of Energy,
grant DE-FG03-92ER40689.}
\footnotetext[23]{Now at Institut Inter-universitaire des hautes Energies (IIHE), CP 230, Universit\'{e} Libre de Bruxelles, 1050 Bruxelles, Belgique}
\footnotetext[24]{Also at Dipartimento di Fisica e Tecnologie Relative, Universit\`a di Palermo, Palermo, Italy.}
\footnotetext[25]{Deceased.}
\footnotetext[26]{Now at SLAC, Stanford, CA 94309, U.S.A}
\footnotetext[27]{Now at INFN Sezione di Roma II, Dipartimento di Fisica, Universit\`a di Roma Tor Vergata, 00133 Roma, Italy.}  
\setlength{\parskip}{\saveparskip}
\setlength{\textheight}{\savetextheight}
\setlength{\topmargin}{\savetopmargin}
\setlength{\textwidth}{\savetextwidth}
\setlength{\oddsidemargin}{\saveoddsidemargin}
\setlength{\topsep}{\savetopsep}
\normalsize
\newpage
\pagestyle{plain}
\setcounter{page}{1}